\documentclass[aps,prd,superscriptaddress,showkeys,showpacs,twocolumn,longbibliography,nofootinbib,floatfix]{revtex4-2}
\usepackage[utf8]{inputenc}
\usepackage{amsmath}
\usepackage{amsfonts}
\usepackage{amsthm}
\usepackage{amssymb}
\usepackage{float}
\usepackage{braket}
\usepackage{graphicx}
\usepackage{xcolor}
\usepackage{url}
\usepackage[colorlinks=true, pdfstartview=FitV, linkcolor=red, citecolor=blue, urlcolor=blue]{hyperref}
\usepackage{slashed}
\usepackage{mathtools}
\usepackage{dsfont}

\usepackage[normalem]{ulem}
\usepackage{fancyhdr}
\usepackage[capitalise]{cleveref}
\usepackage{placeins}

\newcommand{\Tr}{\mathrm{Tr}}
\begin{document}

\title{Thermal nature of confining strings}
\author{Sebastian Grieninger} 
\email{segrie@uw.edu}
\affiliation{InQubator for Quantum Simulation (IQuS), Department of Physics, University of Washington, Seattle, WA 98195}
\affiliation{Department of Physics, University of Washington, Seattle, WA 98195}
\affiliation{Co-design Center for Quantum Advantage (C2QA), Stony Brook University, Stony Brook, New York 11794–3800, USA}
\affiliation{
 Center for Nuclear Theory, Department of Physics and Astronomy,
Stony Brook University, Stony Brook, New York 11794-3800, USA
}

\author{Dmitri E. Kharzeev} 
\email{dmitri.kharzeev@stonybrook.edu}
\affiliation{Co-design Center for Quantum Advantage (C2QA), Stony Brook University, Stony Brook,  New York 11794–3800, USA}
\affiliation{
 Center for Nuclear Theory, Department of Physics and Astronomy,
Stony Brook University, Stony Brook, New York 11794-3800, USA
}
\affiliation{Energy and Photon Sciences Directorate, Condensed Matter and Materials Sciences Division,
Brookhaven National Laboratory, Upton, New York 11973-5000, USA}

\author{Eliana Marroquin} 
\email{eliana.marroquin@stonybrook.edu}
\affiliation{Co-design Center for Quantum Advantage (C2QA), Stony Brook University, Stony Brook, New York 11794–3800, USA}
\affiliation{
 Center for Nuclear Theory, Department of Physics and Astronomy,
Stony Brook University, Stony Brook, New York 11794-3800, USA
}

\begin{abstract}
{We investigate the quantum statistical properties of the confining string connecting a static fermion-antifermion pair in the massive Schwinger model. By analyzing the reduced density matrix of the subsystem located in between the fermion and antifermion, we demonstrate that as the interfermion separation approaches the string-breaking distance, the overlap between the microscopic density matrix and an effective thermal density matrix exhibits a pronounced, narrow peak, approaching unity at the onset of string breaking. This behavior reveals that the confining flux tube evolves toward a genuinely thermal state as the separation between the charges grows, even in the absence of an external heat bath. 
In other words, one cannot tell whether a reduced state of the subsystem arises from a surrounding heat bath or from entanglement with the rest of the system. The entanglement spectrum near the critical string-breaking distance exhibits a rapid transition from the dominance of a single state describing the confining electric string towards a strongly entangled state containing virtual fermion-antifermion pairs.
Our findings establish a quantitative link between confinement, entanglement, and emergent thermality, and suggest that string breaking corresponds to a microscopic thermalization transition within the flux tube.}
\end{abstract}

\maketitle
\thispagestyle{fancy}
\fancyhead{}
\fancyhead[R]{IQuS@UW-21-112, NT@UW-25-15}
\fancyfoot{}
\renewcommand{\headrulewidth}{0pt}
\section{Introduction}

Hadron production in high-energy collisions exhibits thermal features that appear universal and do not depend on the nature of colliding particles and the collision energy. The yields of different hadrons produced in $e^{+}e^{-}$, proton-proton, and heavy-ion collisions are well described by statistical models using a universal temperature \cite{Becattini:1997uf,Andronic_2009}. This apparent thermal behavior is very unlikely to originate from a conventional thermalization process involving rescattering of produced particles. For example, in $e^{+}e^{-}$ annihilation into hadrons, the produced hadrons (and the preceding quarks and gluons) are separated by a space-like interval and cannot undergo multiple final-state interactions. 

Moreover, strong elliptical flow observed in heavy ion collisions at RHIC and LHC points towards a very fast thermalization -- otherwise, the hydrodynamical evolution is too short and is not capable to reproduce the data, see \cite{Schlichting:2019abc} for a review. This early thermalization challenges the conventional models based on kinetic theory \cite{Schlichting:2019abc}. 

An attractive explanation of the observed fast thermalization is \emph{quantum entanglement} -- namely, if an observer performs measurement on a part of a pure quantum system, then tracing over unobserved degrees of freedom can produce a mixed state that locally looks  thermal.

Concrete realizations of this idea include the entanglement produced a rapid pulse of (chromo-)electric field during a high-energy collision \cite{Kharzeev:2005iz,Kharzeev:2006zm,Castorina:2007eb,Florio:2021xvj,Grieninger:2023ehb,Grieninger:2023pyb}. In this case an effective temperature $T$ appears proportional to the inverse duration of the pulse $\tau$, $T \sim 1/\tau$. Another approach \cite{Berges:2017zws,Berges:2017zws} is based on performing the trace over unobserved degrees of freedom in an expanding string described by the Schwinger model. In this case the effective temperature was found to be proportional to the inverse proper time $\tau$, $T = 1/(2\pi\tau)$ \cite{Berges:2017zws,Berges:2017zws}. 

The maximal entanglement emerges from the evolution equations of high energy QCD at small Bjorken $x$ \cite{Kharzeev_2017, Kharzeev:2021nzh}.
It manifests itself via a simple Boltzmann-like relation between the parton structure function $xG(x)$ and the entropy of the produced hadron system $S_h=-\sum p_n \ln p_n$ that is fixed by the independently measured multiplicity distribution $p_n$, $S_h = \ln xG(x)$ \cite{Kharzeev_2017}. This relation was tested in experiment~\cite{Tu:2019ouv,H1:2020zpd}, and found to hold \cite{Kharzeev:2021yyf,Hentschinski:2021aux,Hentschinski:2023izh,Hentschinski:2024gaa}. Maximal entanglement and conformal symmetry \cite{Lipatov:1993yb,Gursoy:2023hge} imply that the small $x$ behavior of the gluon structure function to be $xG(x) \sim x^{-1/3}$ in high-energy QCD~\cite{Grieninger:2025wxg}. The fragmentation of high-energy jets was also found to lead to the maximal entanglement \cite{Datta:2024hpn}. 

The maximal entanglement at a fixed energy corresponds to an effective thermalization, since it yields a reduced density matrix corresponding to the thermal one, with the temperature fixed by the energy. This is in close analogy with  classical statistical physics, where the  Boltzmann postulate of equiprobable micro-state distribution describes, at a fixed energy,  a statistical system in thermal equilibrium.

A microscopic, first-principle, understanding of this emergent thermality within QCD is currently very difficult, as they require real-time Hamiltonian simulations in $(3+1)$ dimensions. Recent studies have therefore turned to lower-dimensional gauge theories where entanglement measures can be computed explicitly. Among these, the massive Schwinger model provides a minimal realization of confinement in $(1+1)$ dimensions, offering analytical tractability and access to large-scale tensor-network simulations.
Indeed, the Schwinger model coupled to external sources was proposed as a way to describe the hadronization of jets by Casher, Kogut and Susskind long time ago \cite{Casher:1974vf}. Analytical solution of this model in the massless fermion limit was found in \cite{Kharzeev:2012re}. 

Numerical real-time simulations of the massive Schwinger model coupled to external sources that describe the jets produced in $e^+e^-$ annihilation (using tensor-networks and exact-diagonalization) were recently performed in \cite{Florio:2025hoc, Florio:2023dke, Florio:2024aix, Janik:2025bbz,Barata:2025hgx}. These studies 
have shown that entanglement entropy (EE) grows rapidly during jet fragmentation. 
The physical content of the Schmidt states (eigenstates of the reduced density matrix) changes as a function of the coupling strength and proper time. While at early times and/or weak coupling the Schmidt eigenstates are close to the Fock quark-antiquark states, at late times and/or strong coupling they represent bound meson states \cite{Florio:2021xvj, Florio:2025hoc}.

The crucially important finding of \cite{Florio:2025hoc} is that at late times the reduced density matrix of the system becomes very close to the thermal one. The resulting effective temperature appears universal, i.e. the comparison of the expectation values of all operators studied to the thermal ones yields the same value of the temperature \cite{Florio:2025hoc}.  These results indicate that entanglement not only underlies hadronization but may also drive the emergence of effective thermal behavior in confining gauge theories. 

A question of fundamental importance is whether this emergent thermal behavior is a consequence of real-time evolution, or is present already within the static confining strings. To address this question, we examine whether entanglement-induced thermality arises in stretched confining strings approaching the break-up point. 

Probing entanglement-induced thermality in stationary settings requires examining the entanglement structure of confining flux tubes. Previously, the tensor-network study \cite{Buyens:2015tea} of the massive Schwinger model characterized confinement and string breaking through static observables and bipartite entanglement entropy, revealing clear signatures of flux tube formation and decay. On this basis, we test whether the reduced density matrix of states within the flux tube looks thermal, or whether the spatial entanglement can generate an effective thermality.

The process of string breaking has been extensively studied within both high-energy and condensed-matter physics.  In spin-chain systems, recent theoretical works have shown that pairs of domain walls bound by a linear potential behave as analogs of particle-antiparticle strings, reproducing key signatures of flux tube formation and decay \cite{Verdel:2019chj,Verdel:2023mmp,Mallick:2024slg,Artiaco:2025qqq}. These concepts are now being tested experimentally in analog quantum simulators, where confinement dynamics have been observed in trapped-ion chains, Rydberg atom arrays, and superconducting qubit architectures \cite{Alexandrou:2025vaj,Crippa:2024hso,Liu:2024lut,De:2024smi,Surace:2024bht}. 
Tensor-network and hybrid quantum–classical approaches have captured both static and dynamical aspects of string formation and breaking in one-dimensional $\mathbb{Z}_2$ gauge models \cite{Alexandrou:2025vaj}, while recent extensions to two spatial dimensions demonstrate that confinement and screening persist in experimentally accessible geometries \cite{Cochran:2024rwe,Gonzalez-Cuadra:2024xul,Borla:2025gfs,Cataldi:2025cyo,Xu:2025abo,DiMarcantonio:2025cmf}. 
Together, these developments establish a unified framework in which confinement and string breaking can be probed across high-energy, condensed-matter, and quantum-information platforms, setting the stage for a microscopic exploration of entanglement and thermality in static flux tubes.

In this work, we investigate the emergence of thermal features in a static confining flux tube, focusing on the ground state of the massive Schwinger model with two external charges. This setup enables us to isolate the role of spatial entanglement in a stationary configuration. By varying the separation between the static charges, we probe the evolution of local observables, the entanglement entropy, and the entanglement spectrum of a subsystem. We further quantify the degree of thermalization by computing the overlap between the reduced density matrix of this subsystem and a thermal ensemble with the same energy.

Our results show that as the separation approaches the string-breaking distance, the entanglement entropy peaks, the entanglement-spectrum gap closes, and the reduced density matrix becomes thermal. These signatures identify string breaking as a microscopic thermalization transition within the confining flux tube, establishing a stationary realization of entanglement-induced thermality previously associated with dynamical evolution. This provides a direct mechanism by which confinement and screening generate an effective thermal behavior, a connection that may underlie the universal thermal patterns observed in QCD hadronization.

Note that the emergence of thermality and temperature is very natural in the holographic picture of a Schwinger pair~\cite{Grieninger:2023ehb,Grieninger:2023pyb,Jensen:2013ora,Sonner:2013mba}. Since the fermion and anti-fermion are spatially separated, the minimal surface connecting the two (in the extra dimension) exhibits a world-sheet horizon which is associated with a temperature. Moreover, the thermal properties of bound states of the Schwinger model were discussed in~\cite{Asadi:2022vbl,Asadi:2023bat}.
\bigskip

The paper is organized as follows. In Section \ref{model}, we introduce the lattice formulation of the massive Schwinger model and describe the implementation of the static external charges. Section \ref{secresults} presents our numerical results. We begin by determining the ground-state potential and characteristic energy scales associated with the string breaking, followed by an analysis of the spatial dependence of the energy density, pressure and charge distribution. We then examine the chiral condensate and electric-field energy, which complement the previous observables and further elucidate the internal structure of the flux tube. Next, we analyze the bipartite entanglement entropy and the entanglement spectrum as probes of confinement and screening, capturing the buildup of correlations within the flux tube and subsequent decay upon string breaking. We then extract an effective temperature by comparing reduced density matrices to thermal ensembles, establishing a quantitative link between spatial entanglement and emergent thermality. Finally, in Section \ref{sec_discussion}, we summarize and discuss our results and their implications.

\section{The model}\label{model}
To analyze the quantum statistical properties of string breaking, we will use a model that shares many properties with QCD and is amenable to current quantum simulations -- namely, the massive Schwinger model  \cite{Schwinger2}. The model is quite simple, but it  exhibits the key features of confinement and chiral symmetry breaking. 

The Schwinger model describes a single flavor Dirac fermion $\psi$ of mass $m$ and charge $g$ coupled to a U(1) gauge field $A_\mu$ with field strength tensor $F_{\mu\nu} = \partial_\mu A_\nu - \partial_\nu A_\mu$.
In temporal gauge this reduces to a single component $F_{01}=E$ describing an electric field; there are no propagating photons. 

The corresponding Hamiltonian is given by 
\begin{align}
    H = \int dx &\left( \frac{1}{2}E^2 + \bar{\psi}( -i\gamma^\mu \partial_\mu + g\gamma^1 A_1 +m )\psi \right) 
\nonumber \\ &+ \int dx\ j_{ext}^1 (x) A_1
\label{continuum_SM}
\end{align}
where $j_{\textrm{ext}}^1(x)$ is an external source. We model the string by introducing the static sources describing a fermion and an antifermion (with an electric field in between them). 

We study the process of string breaking by discretizing the theory on a spatial lattice, mapping the continuum theory onto a finite-dimensional Hilbert space.   We adopt the Hamiltonian lattice formulation, introduced by Kogut and Susskind \cite{kogutsuss}, in which the space is discretized while time remains continuous. 

The fermionic fields are represented by staggered lattice operators $\chi_n$ and $\chi_n^\dagger$, staggering the upper and lower components on even and odd sites, respectively, while the gauge degrees of freedom ``live" on links connecting the  neighboring sites. The electric field on each link $L_n$ is conjugate to the compact link operator $U_n =e^{iA_n}$, satisfying the relation $[L_n,U_m]=\delta_{nm}U_n$. The lattice Hamiltonian for the Schwinger model \eqref{continuum_SM}:
\begin{align}
    H =  - \frac{i}{2a}& \sum_{n=1}^{N-1} ( \chi_n^\dagger U_n \ \chi_{n+1} - \chi_{n+1}^\dagger U_n^\dagger \chi_n)  + m \sum_{n=1}^{N} \chi_n^\dagger \chi_n \nonumber \\ & + \frac{g^2a}{2} \sum_{n=1}^{N-1} (L_n +L_{\text{ext},n})^2
    \label{lattice_SM}
\end{align}
Here, $N$ is the even number of lattice sites, $a$ the lattice spacing, and $g$ the coupling constant.  Using the open boundary conditions, we can integrate out the gauge fields by imposing  Gauss law as a constraint on every site,
\begin{equation}
    L_n - L_{n-1} =Q_n  \quad \textrm{and } \quad Q_n \equiv \chi_n^\dagger \chi_n - \frac{1 - (-1)^n}{2}, 
    \label{gauss}
\end{equation}
where $Q_n$ is the local charge operator. Indeed, by using a gauge transformation to fix $U_n= 1 $, the Gauss law \eqref{gauss} can be solved to express the electric field recursively as
\begin{equation}
    L_n = L_0 + \sum_{k=1}^n Q_k.
\end{equation}
Substituting this relation back into the Hamiltonian \eqref{lattice_SM} removes the explicit gauge degrees of freedom and yields a purely fermionic formulation.

In the presence of external static charges, the total electric field acquires an additional contribution $L_{\textrm{ext,}n}$. For a string of length $d$ centered at site $N/2$ and created by a fermion and an antifermion, the external field is implemented as
\DeclarePairedDelimiter\abs{\lvert}{\rvert}%
\begin{equation}
    L_{\textrm{ext,}n}(d) = - \Theta\!  \left( \frac{d/a - 1}{2} - \abs*{ n -  \frac{N}{2}} +\delta\right)
\end{equation}
where $\Theta$ is the Heaviside step function and $\delta$ is a small number. In the absence of dynamical screening, this term introduces a region of a constant electric field between the two static charges, representing the confining flux tube connecting them. 

\begin{figure}[h] 
\centering
     \includegraphics[width=1\linewidth]{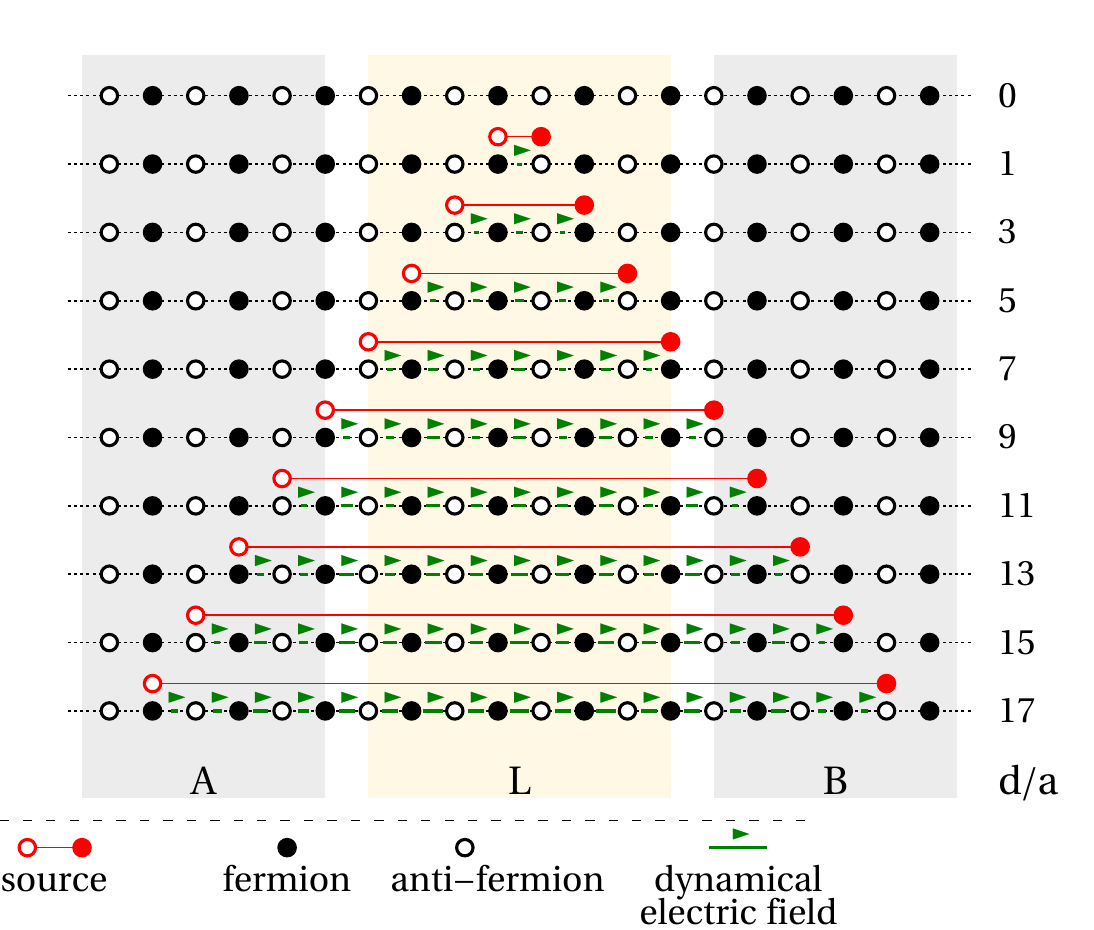}\caption{Cartoon of our setup using $N=20$ staggered sites as an example. The yellow shaded region exemplifies a centered subsystem of size $L=8$ in the middle of the chain. The first row shows the vacuum state without external source which is our reference state. The second row contains the external sources at separation $d=1\,a$, the third at $d=3\,a$, etc.}
     \label{setup_fig}
\end{figure}

To map the fermionic degrees of freedom to spin variables, we apply the Jordan-Wigner transformation \cite{PhysRevD.16.3031}, which expresses the staggered fermion operators in terms of Pauli matrices:
\begin{equation}
    \chi_n = \frac{X_n - iY_n}{2} \prod_{m=1}^{n-1} Z_m, 
\end{equation}
where $X_n$, $Y_n$, and $Z_n$ denote the Pauli operators acting in the local spin-1/2 Hilbert space of site n. Substituting this relation into the lattice Hamiltonian results in the final expression
\begin{align}
    H (d)=  \frac{1}{4a}&  \sum_{n=1}^{N-1}\left( X_n X_{n+1} + Y_n Y_{n+1} \right)+  \frac{m}{2} \sum_{n=1}^{N} (-1)^n  Z_n \nonumber \\ +&\frac{g^2a}{2} \sum_{n=1}^{N-1} (L_n +L_{\text{ext},n}(d))^2  .
\end{align}
Our setup including the external charges is visualized in Figure~\ref{setup_fig}.

\section{Results} \label{secresults}
In the following, all observables are shown as functions of the dimensionless separation $d\cdot M_s$ between the two external static charges. Unless otherwise stated, all results correspond to $N=220,\, m=0.045\ a^{-1}$, and $g=0.09\ a^{-1}$.
Throughout this work, we set $a=1$ and express temperatures and other dimensionful quantities in physical units defined by the mass of the first excited state of the Schwinger model, $M_s$, corresponding to the (pseudo)scalar meson. Our numerical results are obtained using the \textit{ITensor} software library~\cite{Fishman_2022} in \textit{Julia}~\cite{bezanson2015juliafreshapproachnumerical}.

\subsection{String breaking and energy scales}
\begin{figure}[ht!]
     \centering     \includegraphics[width=0.7\linewidth]{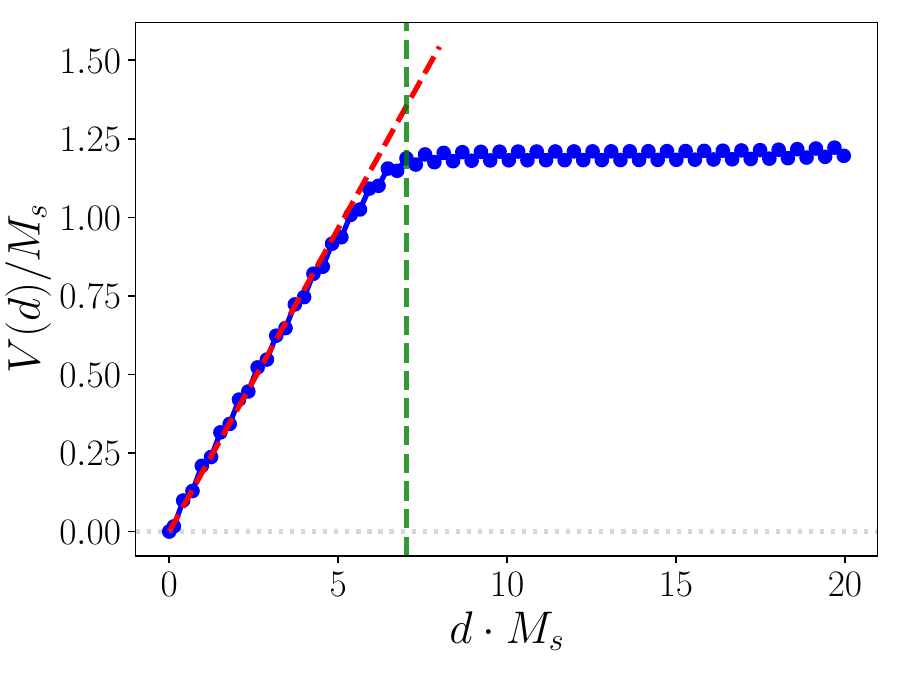}
     \includegraphics[width=0.7\linewidth]{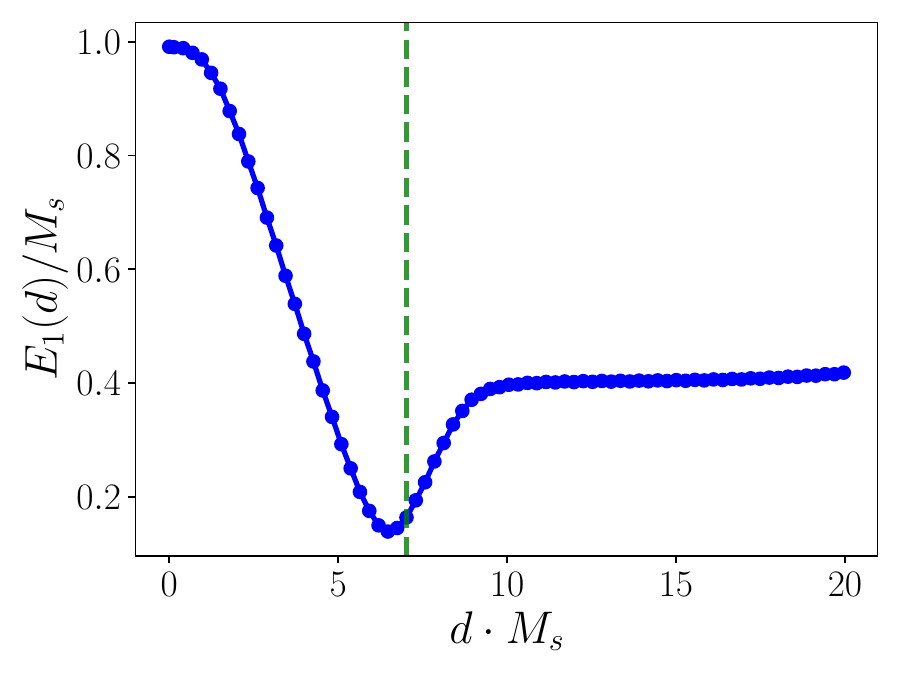}  
     \caption{Ground-state potential $V(d)$ (top) and normalized mass gap of the first excited state $E_1(d)/M_s$ (bottom) as functions of the separation $d\cdot M_s$ between the external static charges ($M_s\equiv E_1(0)$ is the mass of the (pseudo)scalar boson). The potential rises linearly at short distances, reflecting confinement. We define the critical separation $d_c$ (shown by the vertical dashed line) as the distance at which the derivative of the potential drops by 80\% of its plateau value; this gives the critical distance of $d_c\cdot M_s \simeq 7$. At this distance, the energy stored in the flux tube reaches the threshold for pair creation. The linear increase is elucidated by the red dashed line which is a linear fit with $V/M_s\sim 0.193\, d\cdot M_s$. The plateau level is $V/M_s\sim 1.09$. The mass gap exhibits a minimum at the same distance, providing an independent signature of string breaking.}
     \label{stringbreaking}
\end{figure}

We first determine the characteristic energy scales of the system by computing the ground-state potential and the lowest excitation gap. Figure \ref{stringbreaking} displays these quantities. The potential rises linearly at short distances, reflecting the formation of a uniform electric flux tube with a constant string tension. Around $d_c\cdot M_s  \simeq 7$, the potential saturates, indicating that the energy stored in the flux tube equals the threshold for dynamical fermion anti-fermion pair creation. For larger separations, the string breaks and the system rearranges into two screened meson-like states, leading to a constant asymptotic energy. 

The normalized gap of the first excited state  $E_1(d)/M_s$ exhibits a minimum at approximately the same distance where the potential flattens, 
indicating an increase in the density of states that accompanies the string breaking. The minimum in the excitation gap thus 
can provide an independent signature of string breaking. The critical separation $d_c$ that separates the linearly confining regime from the screened phase serves as a reference scale for the analysis below.

\subsection{Energy density and charge distribution}
\begin{figure}[ht!]
     \centering
     \includegraphics[width=0.8\linewidth]{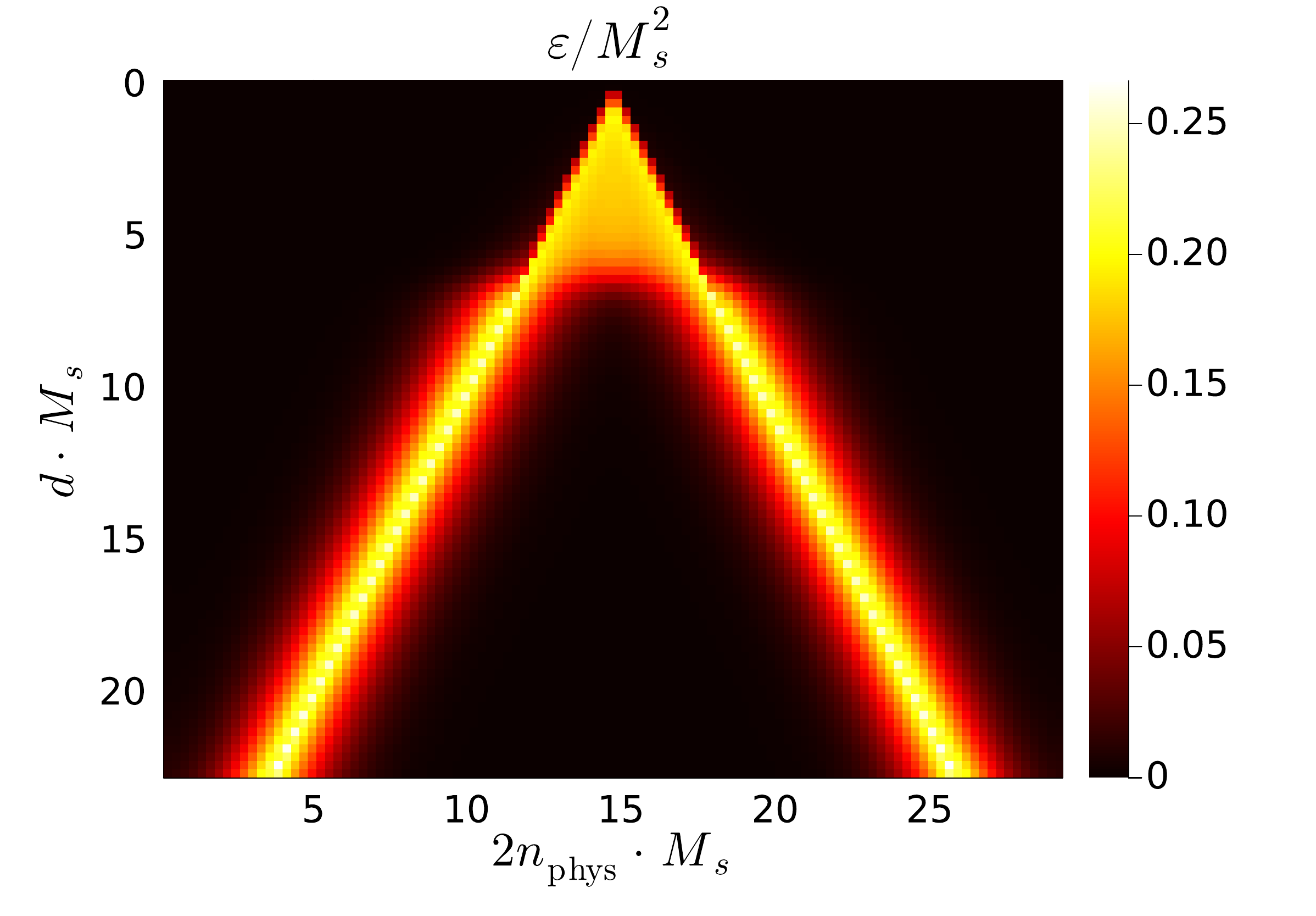}
     \includegraphics[width=0.8\linewidth]{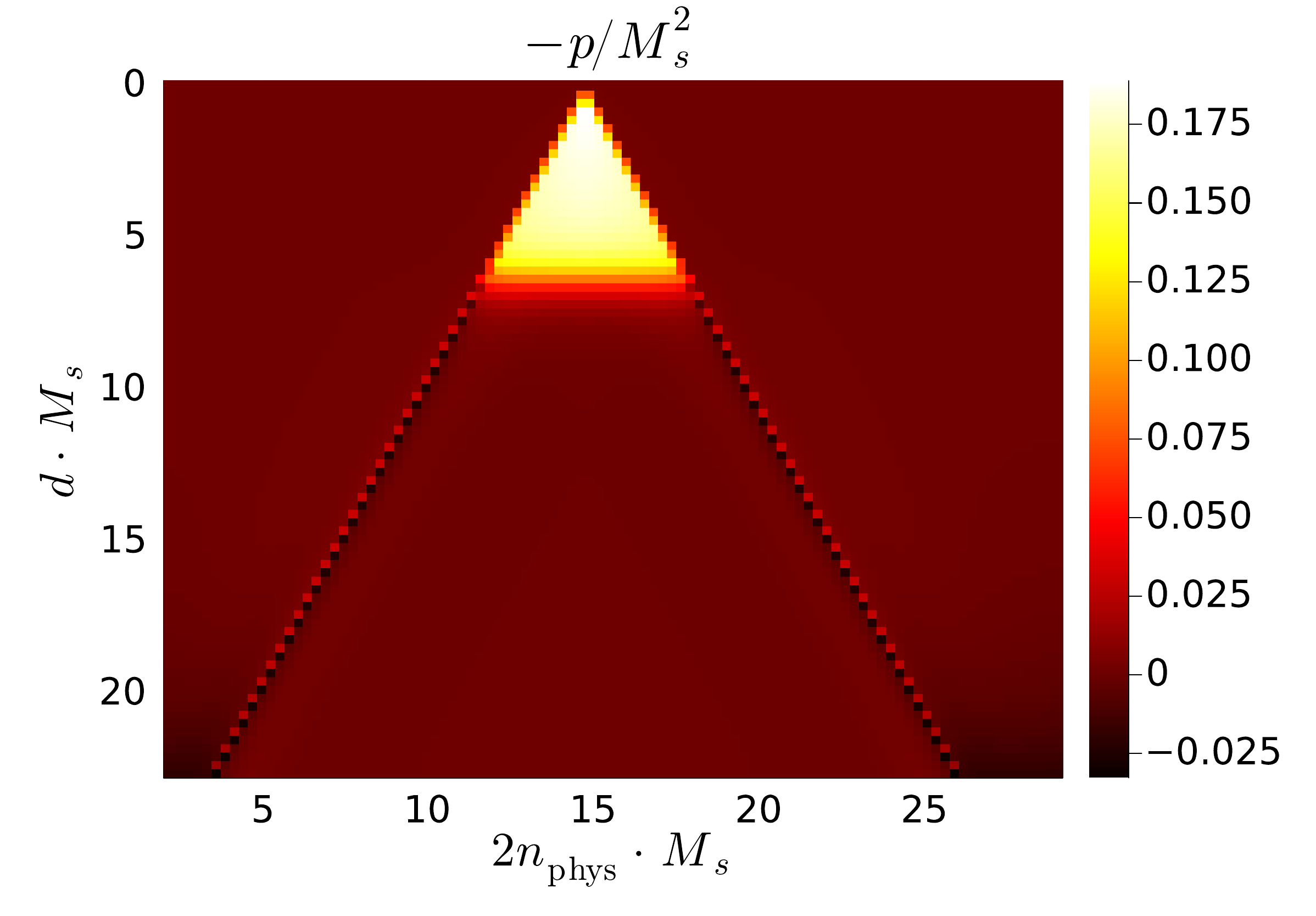}
     \includegraphics[width=0.8\linewidth]{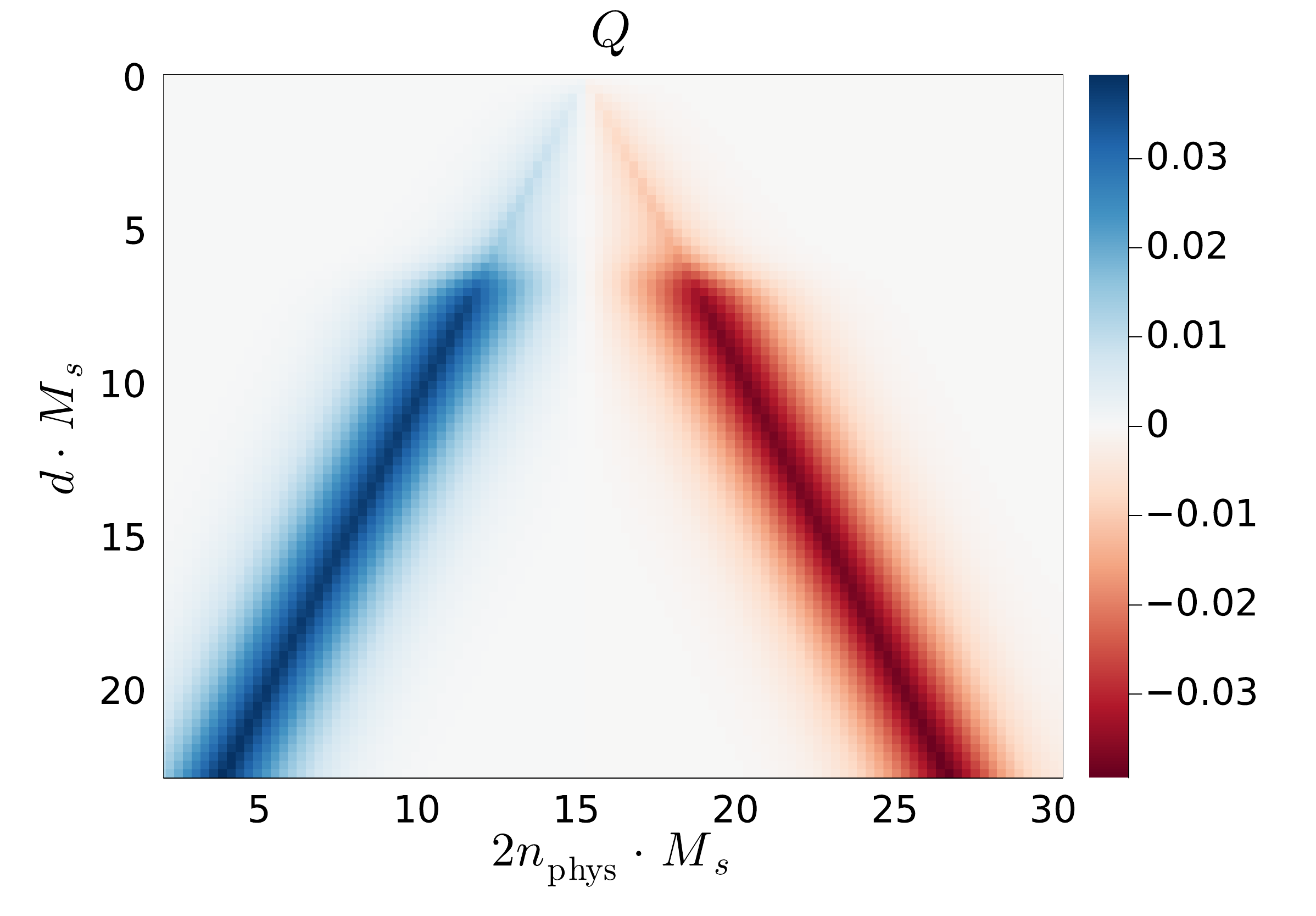}     
     \caption{Spatial profiles of the ground-state energy density (top), pressure (middle), and charge density (bottom) as functions of the separation $d\cdot M_s$ between the external charges (vertical axis) and the spatial coordinate along the lattice spin chain (horizontal axis). See Fig. \ref{setup_fig} for the setup. The energy density illustrates the formation and subsequent decay of the confining flux tube, while the charge distribution shows the emergence of a dynamical fermion–antifermion pair that screens the external sources. At large separations, the flux tube disappears and two neutral meson-like states remain.}
     \label{spatial}
\end{figure}

The spatial structure of the ground-state energy density, pressure, and charge density further illustrates the transition between the confined and screened regimes, as shown in Figure \ref{spatial}. 
In the ground state, the energy density and pressure are given by the expectation values of the $T^{00}$ and $T^{11}$ components of the energy momentum tensor, respectively (see for example~\cite{Florio:2025hoc}).

At short separations, the energy density forms a single localized region between the sources, corresponding to a uniform electric flux tube, see Figure \ref{spatial}, top. As the separation increases, the stored energy  makes it energetically favorable to create the dynamical fermion-antifermion pair from the vacuum. Note that this fall-off is not abrupt, because in massive Schwinger model the produced mesons interact, and this interaction affects the energy density near the critical distance.

The pressure (see Figure \ref{spatial}, middle) exhibits a similar behavior: at smaller separations, the unscreened electric field creates a negative pressure, which gradually disappears near the critical string-breaking distance. 

The charge density, $q_n=(\langle Z_n\rangle+(-1)^n)/(2a)$,  behaves accordingly: it is initially localized near the external sources with a neutral interior, but new peaks of opposite sign emerge near the center around $d_c\cdot M_s \simeq 7$, indicating the pair creation and subsequent screening, see Figure \ref{spatial}, bottom. At large separations, the flux tube disappears, leaving two neutral meson-like bound states.

\subsection{Chiral condensate and electric energy} 
\begin{figure}[ht!]
     \centering
     \includegraphics[width=0.8\linewidth]{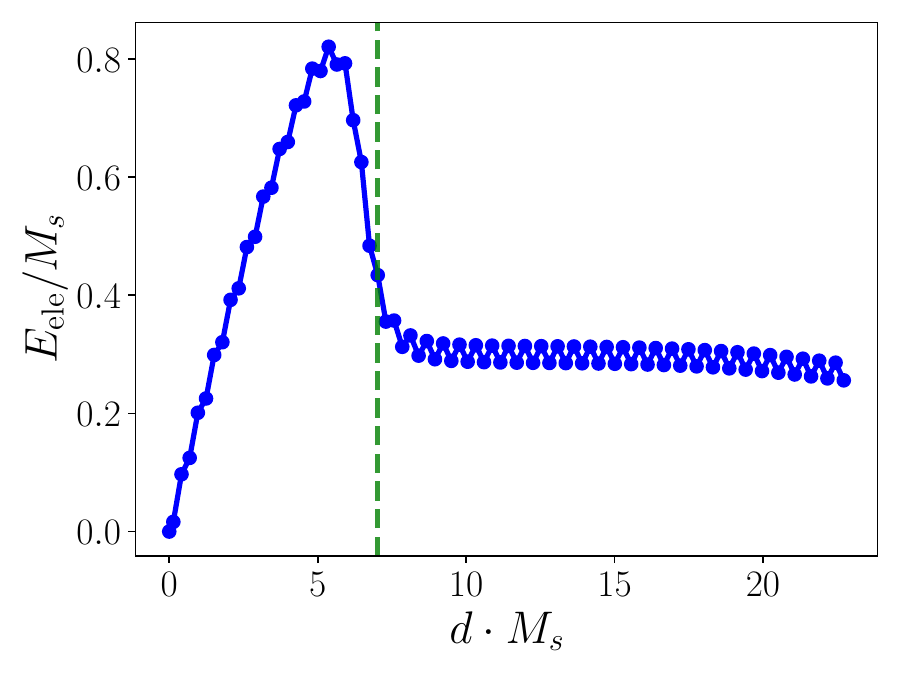}
     \includegraphics[width=0.8\linewidth]{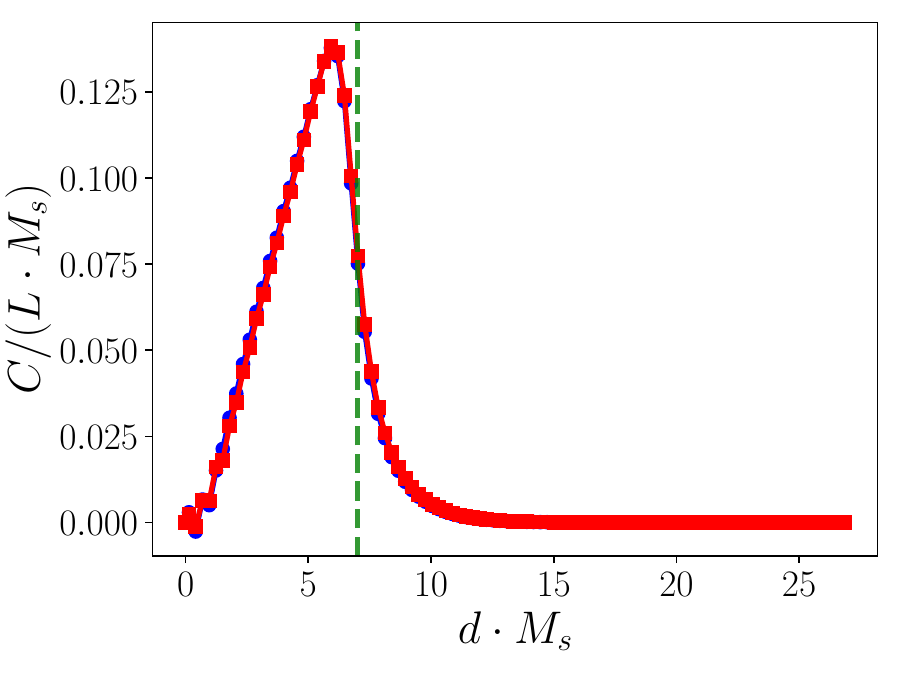}   
     \caption{Electric-field energy (top) and spatially averaged condensate (bottom) of the central 8 (blue circles) and 12 (red squares) sites, respectively, shown relative to their vacuum values. The electric energy increases with the separation between the static charges, reaching a maximum near the critical distance $d_c  M_s \simeq 7$, where the system becomes unstable to pair creation. The chiral condensate exhibits the opposite trend, being suppressed inside the flux tube and attaining maximal reduction at the same distance, signaling partial restoration of chiral symmetry within the confined region.}
    \label{obs3}
\end{figure}

We resolve the internal structure of the flux tube by computing a set of local observables measured relative to the ground state in the absence of external charges, as shown in Figure \ref{obs3}. 
The electric-field energy, defined as $E_\text{ele}\equiv ag^2/2\,\sum_{n=1}^{N-1}\langle (L_n+L_{n,\text{ext}}(d))^2\rangle$, increases with energy separation, reflecting the growing field strength between the static charges, and peaks near $d_c$, where the system becomes unstable against pair creation. Beyond this point, screening by the produced charges reduces the stored energy.

The spatially averaged chiral condensate $C_n=(-1)^n\langle Z_n\rangle/(2a)$, evaluated over the central sites, shows the opposite behavior. It is suppressed inside the flux tube as the strong field polarizes the vacuum, reaching maximal reduction near the same critical distance. The correlation between field energy and condensate suppression indicates a local tendency toward chiral-symmetry restoration inside the flux region.

\subsection{Entanglement entropy}

The entanglement entropy (EE) for which two types of partitions are considered: a subsystem of length $L$ centered in the middle of the lattice, and a half-chain bipartition separating the system into left and right halves, see Appendix \ref{app:sre} for details. The first probes the spatial extent of entanglement along the flux tube, while the second captures the global buildup of correlations as the external charges are separated. 

\begin{figure}[ht!]
    \centering
    \includegraphics[width=0.8\linewidth]{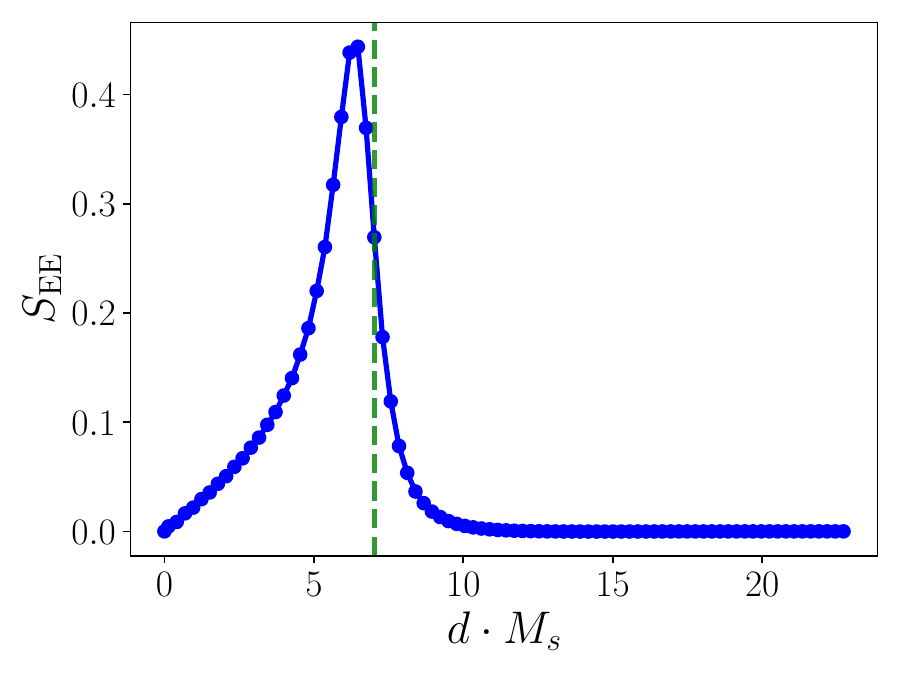}
    \includegraphics[width=0.8\linewidth]{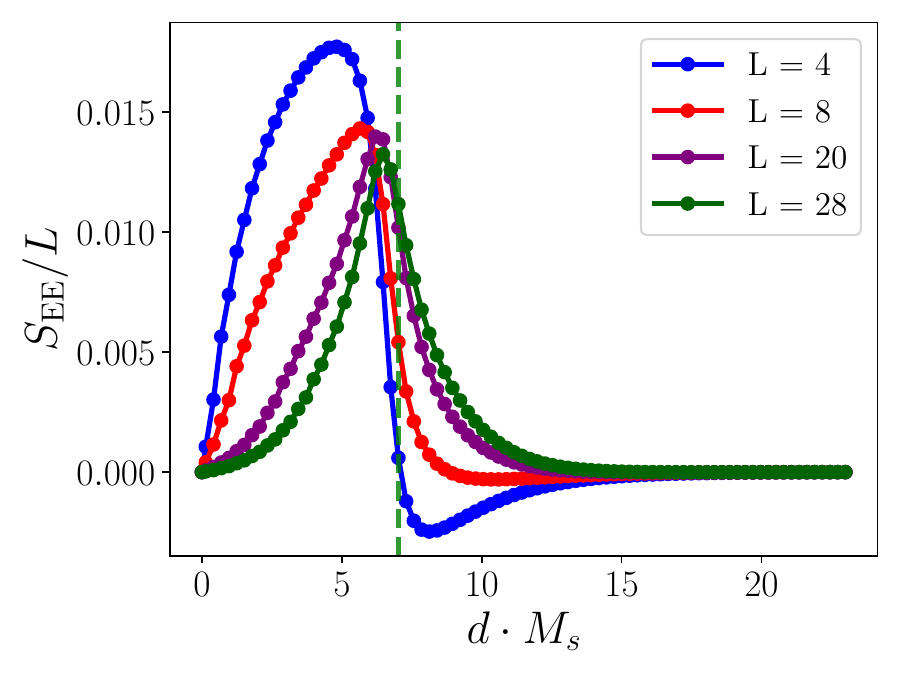}    
    \caption{Entanglement entropy as a function of the charge separation $d \cdot M_s$ for different partitions of the system. \textbf{Top:} Half-chain entropy capturing the total entanglement between the left and right halves of the lattice. \textbf{Bottom:} Entanglement entropy of centered subsystems of length $L$, measured relative to the vacuum -- hence the region of negative $S_{EE}$. }
    \label{entropy}
\end{figure}

The half-chain entropy, shown in the top panel of Figure \ref{entropy}, displays a striking non-monotonic behavior: it rises sharply with separation, reaches a peak near the critical distance, and then returns toward its vacuum value at large $d$.   This peak provides a clear signature of confinement and its decay, consistent with the tensor-network results of \cite{Buyens:2015tea}, where the EE was identified as a sensitive probe of string formation and screening.

Figure \ref{entropy} (bottom) shows the EE for various subsystem sizes $L$, measured relative to the vacuum. For small separations, the entanglement is weak, as the system remains close to a product state dominated by the external sources. As the distance $d$ increases, the EE grows, reflecting the buildup of quantum correlations within the confining flux tube.  
Beyond  $d_c$, the entanglement rapidly decreases as the flux tube fragments into two screened meson-like states. Note that we show the difference between the EE and its value in the vacuum -- so the negative value of $S_{EE}$ observed for the smallest size $L=4$ of the subsystem indicates that the entanglement entropy becomes smaller than its vacuum value. This may be due to the fact that for the smallest subsystem the produced mesons that are close to each other screen the entanglement correlations and make them smaller than in the vacuum. 

We have also investigated the EE within different charge sectors (symmetry-resolved EE). We have observed that the charge-2 sector possesses even larger EE than the charge-0 one, Appendix \ref{app:sre}. The weight of this charge-2 sector reaches its peak near the critical string-breaking distance, when the two mesons are formed. For more details and results for other symmetry-resolved sectors, see Appendix \ref{app:sre}.

\subsection{Entanglement spectrum}
\begin{figure}[ht!]
     \centering
     \includegraphics[width=0.8\linewidth]{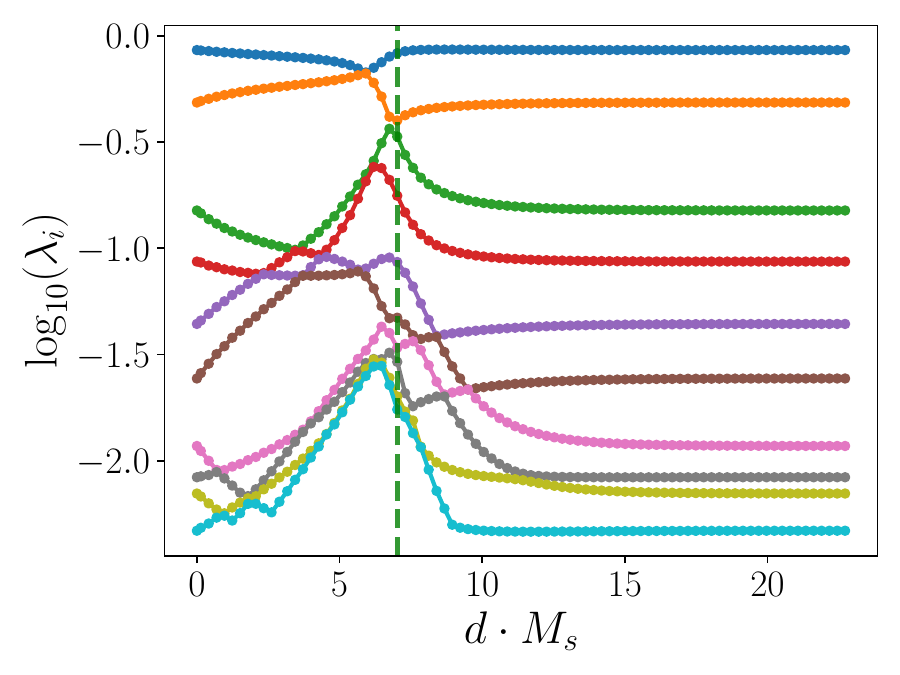}   
          \includegraphics[width=0.8\linewidth]{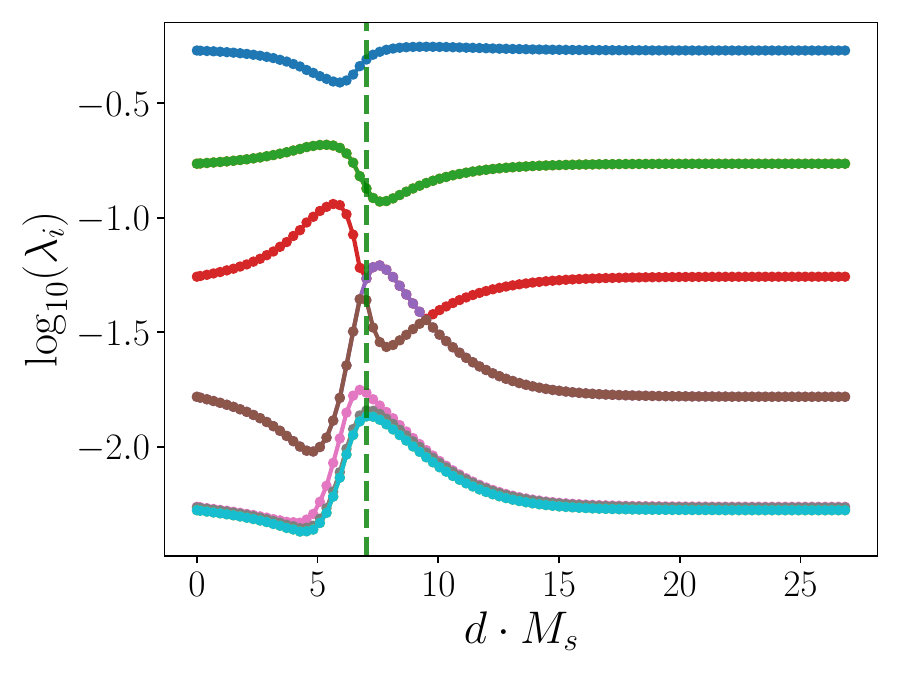}   
     \caption{\textbf{Top:} entanglement spectrum for half of the system which yields the bipartite entanglement entropy shown in Fig. \ref{entropy}.
     \textbf{Bottom:} Entanglement spectrum of the reduced density matrix $\rho_A$ for a centered $L=32$ subsystem, shown as the logarithm of the ten largest eigenvalues versus the separation between the external charges $d \cdot M_s$. }
     \label{spectrum}
     \end{figure}

The entanglement spectrum of the reduced density matrix $\rho_A$, shown in Figure \ref{spectrum}, captures the dependence of quantum correlations within the flux tube region on the separation between the external charges. The figure presents the logarithm of the ten largest eigenvalues of $\rho_A$ as a function of the separation between the external charges. As $d$ approaches the string-breaking distance $d_c\cdot M_s \sim 7$, the eigenvalues cluster and the spectral gap closes, indicating that several Schmidt states contribute with comparable weight. This redistribution of spectral weight signals enhanced quantum mixing and an approach toward maximal entanglement within the subsystem, consistent with the peak in entanglement entropy at the same separation.

Near the critical distance $d_c$, there is a level crossing for the two most important eigenstates, see Fig. \ref{spectrum}. We interpret it as follows: at distances $d<d_c$, the dominant state corresponds to the external charges connected by a confining electric string, with no additional fermion-antifermion pairs. Near the critical distance $d_c$, the admixtures of higher Fock states with additional fermion-antifermion pairs become important. At distances $d>d_c$, the dominant state corresponds to the external charges screened by the produced fermion and antifermion, a two-meson state. At the distance at which the level crossing takes place, we have a maximal entanglement of the two states: external static charges connected by the confining string, and the two-meson state describing the screened external charges.

Past the critical distance $d_c$, the spectrum does not exhibit any dependence on the distance, and $\rho_A$ approaches a  mixed state independent of the positions of the sources. 
Near the critical distance $d_c$, the eigenvalues of the Schmidt states become close to each other, signaling an approach to maximal entanglement. This indicates that the subsystem loses sensitivity to the microscopic structure of the global wavefunction. Such behavior parallels that of a thermal ensemble, where local observables are governed by statistical rather than coherent correlations. This transformation in the spectral structure results in the maximum of the entanglement entropy and provides a signature of local thermalization within the flux tube, a conclusion further supported by the thermal-overlap analysis presented in the next section.

\subsection{Effective temperature from thermal overlaps}

We quantify the degree of thermalization of the flux tube subsystem by comparing its reduced density matrix $\rho_A(d)$ to thermal density matrices $\rho_\beta$ of the static Hamiltonian at temperature $T = 1/\beta$. 
Following the prescription introduced in \cite{Florio:2021xvj}, we first obtain $\rho_A(d)$ from the ground state of the system with static sources separated by a distance $d$, tracing out all sites outside a centered region of length $L$. 
Separately, we construct thermal states for a system of length $N_2>L$ without external sources and compute their reduced thermal density matrices $\rho_\beta$ of a subsystem of the same size $L$. 
The comparison between $\rho_A$ and $\rho_\beta$ then determines the effective temperature $T(d)$. 

In principle, this can be achieved by maximizing the fidelity or minimizing the trace distance between $\rho_A$ and $\rho_\beta$. 
However, because these quantities scale unfavorably with subsystem size, we instead employ computationally efficient proxies: the normalized overlap $f(\rho_A,\rho_\beta)$ and the Hilbert-Schmidt distance $D_{\mathrm{HS}}(\rho_A,\rho_\beta)$, defined as
\begin{align}
     f(\rho_A,\rho_\beta) &\equiv \frac{\Tr (\rho_A \rho_\beta)}{\sqrt{\Tr\,\rho_A^2 \, \Tr\,\rho_\beta^2}}\,, \label{overlap_definition}\\
      D_\text{HS}(\rho_A,\rho_\beta)&=\sqrt{\Tr[(\rho_A-\rho_\beta)^2]}\,.
\end{align}
 \begin{figure*}[ht!]
     \centering
     \includegraphics[width=0.32\linewidth]{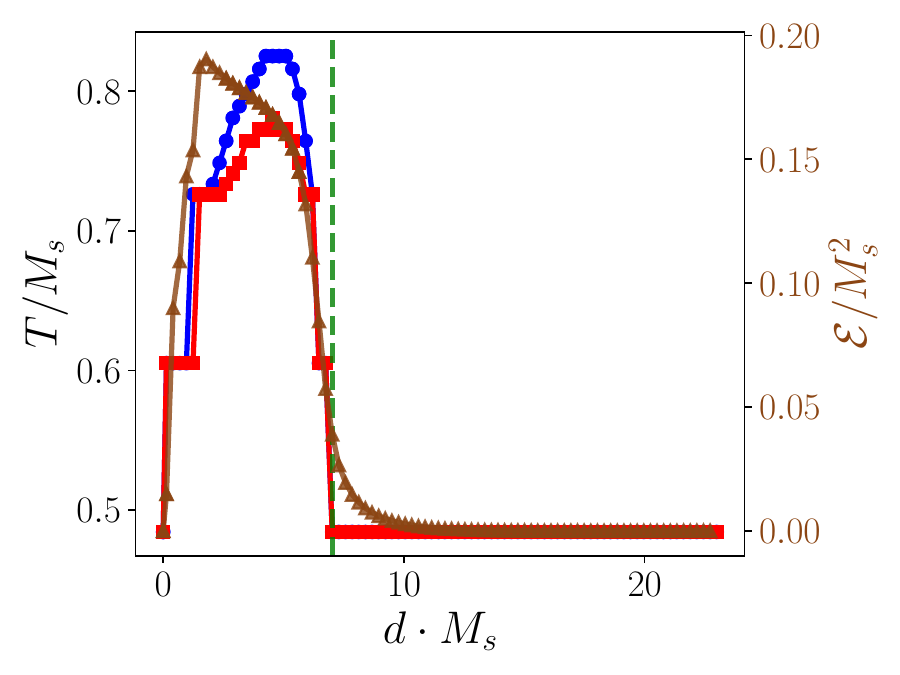}     \includegraphics[width=0.32\linewidth]{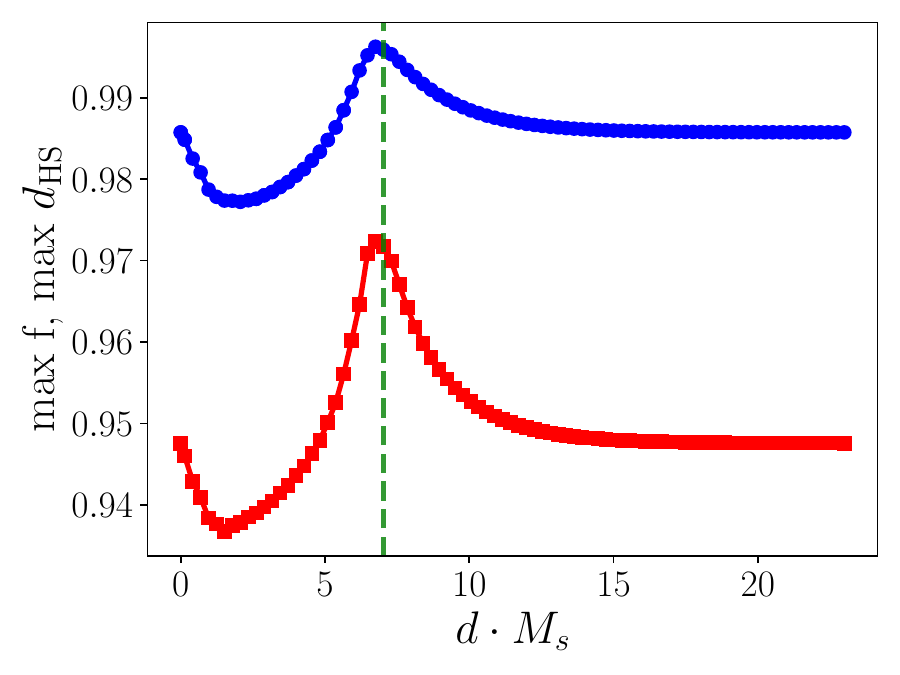}
     \includegraphics[width=0.32\linewidth]{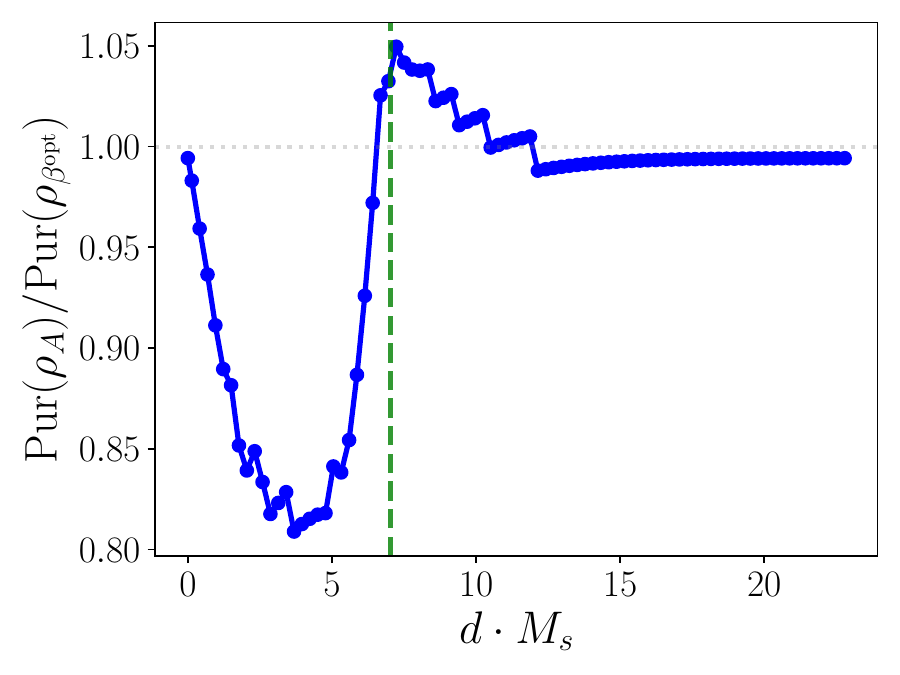} 
     \caption{Subsystem: $L=8$.  Blue dots: Temperature extracted from the maximal normalized overlap (\textbf{left}) and the corresponding maximal normalized overlap as a function of $d\cdot M_s$ (\textbf{middle}). Red squares: Temperature from minimal Hilbert-Schmidt distance (\textbf{left}) and the corresponding complement of Hilbert-Schmidt distance at its maximum value as a function of $d\cdot M_s$ (\textbf{middle}). For comparison, the energy density is added (brown triangles). The \textbf{right} plot shows the ratio of the purity of the reduced density matrix to the purity of the thermal density matrix at maximal normalized overlap.}
     \label{overlap8}
     \end{figure*}

     \begin{figure*}[ht!]
     \centering
     \includegraphics[width=0.32\linewidth]{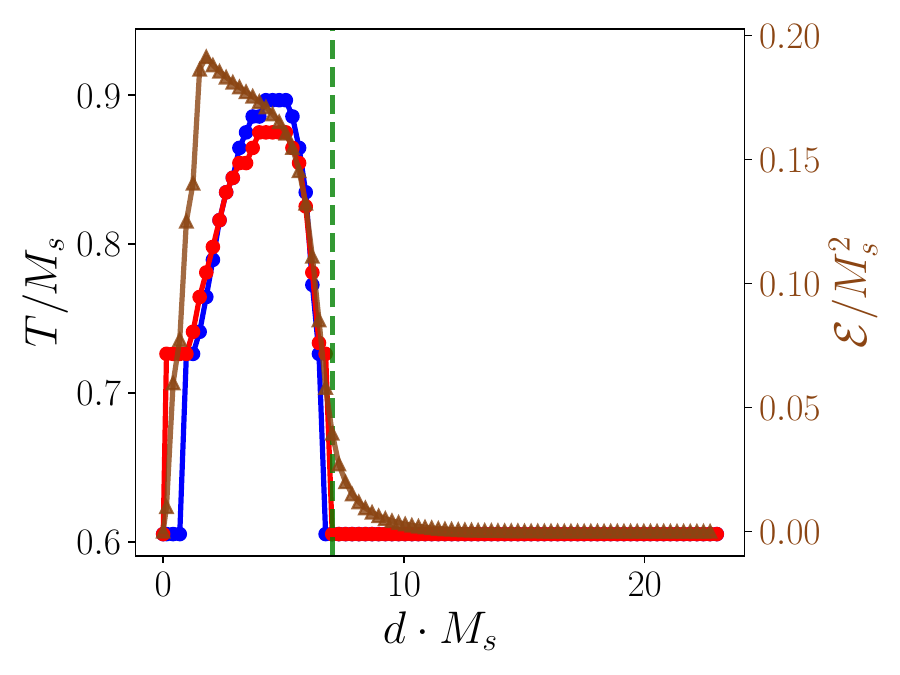}     \includegraphics[width=0.32\linewidth]{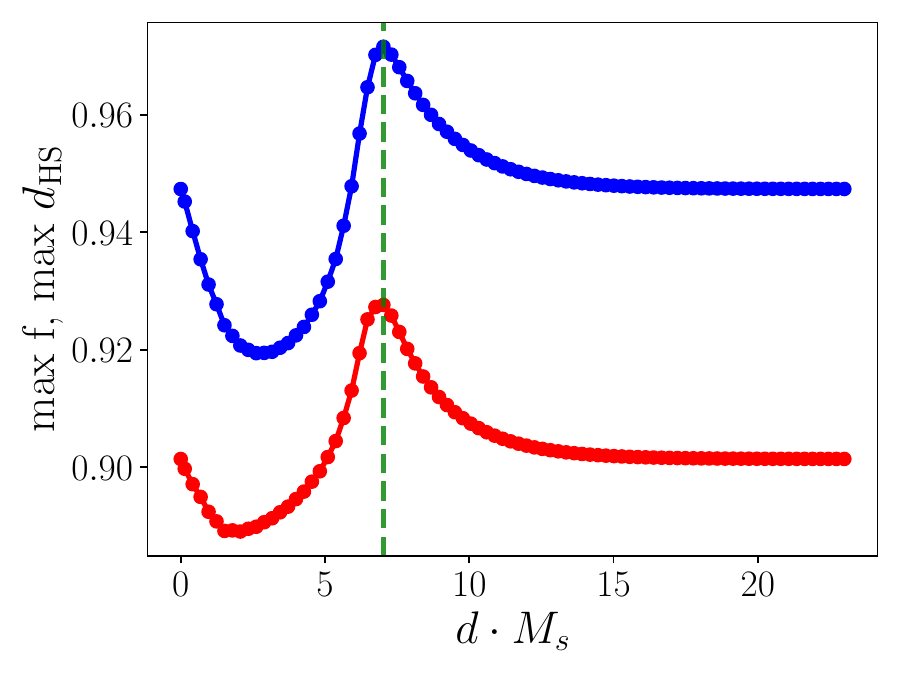}
     \includegraphics[width=0.32\linewidth]{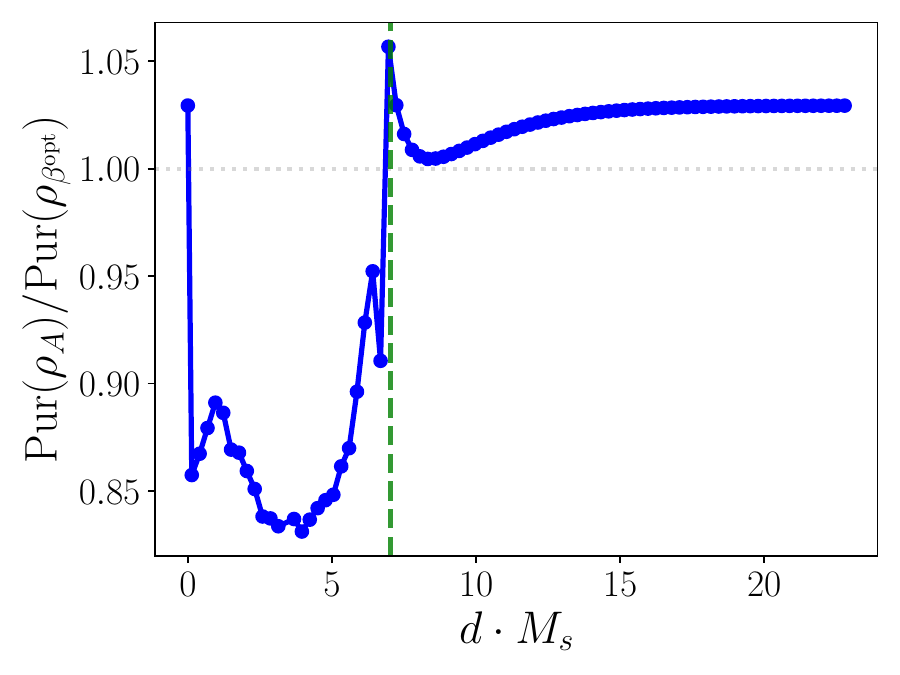}    
     \caption{Subsystem: $L=12$. Blue dots: Temperature extracted from the maximal normalized overlap (\textbf{left}) and the corresponding maximal normalized overlap as a function of $d\cdot M_s$ (\textbf{middle}). Red squares: Temperature from minimal Hilbert-Schmidt distance (\textbf{left}) and the corresponding complement of Hilbert-Schmidt distance at its maximum value as a function of $d\cdot M_s$ (\textbf{middle}). For comparison, the energy density is added (brown triangles). The \textbf{right} plot shows the ratio of the purity of the reduced density matrix to the purity of the thermal density matrix at maximal normalized overlap.}
     \label{overlap12}
     \end{figure*}

For convenience, we also define the complement of the Hilbert–Schmidt distance,
\begin{equation}
    d_\text{HS}(\rho_A,\rho_\beta)=1-D_\text{HS}(\rho_A,\rho_\beta),
    \label{small_d}
\end{equation}
so that in both cases the effective temperature is identified by maximizing \eqref{overlap_definition} and \eqref{small_d}. 
Thermal wave functions are generated using the purification method (see Appendix D of \cite{Florio:2025hoc} for details).

The behavior of these overlap measures connects directly to the entanglement-spectrum analysis of the previous section. 
A large normalized overlap or a small Hilbert-Schmidt distance indicates that the reduced subsystem is nearly indistinguishable from a thermal ensemble, thereby confirming that the flattened entanglement spectrum indeed corresponds to an effectively thermal mixed state. 
Figures \ref{overlap8} and \ref{overlap12} display the extracted temperature and overlap measures for subsystems of length $L=8$ and $L=12$, respectively. 

In both cases, the effective temperature increases with the separation between the external charges and reaches a pronounced maximum near the critical distance $d_c \cdot M_s \simeq  7$, where the string breaks. Note that as soon as the string breaks, the temperature reaches its vacuum value, even though the transition for the energy density is more smooth (presumably due to the residual interaction between the produced mesons). It is important to keep in mind that an effective temperature of the quantum vacuum is different from zero \cite{Florio:2025hoc}. 

The overlap measures shown in the middle panels peak in the same region, demonstrating that the similarity between $\rho_A$ and $\rho_\beta$ is strongest near the critical distance $d_c$. For subsystem $L=8$ the overlap is maximum right before $d_c$ and exactly at $d_c$ for $L=12$.
For the larger subsystem, the temperature distribution becomes broader and the normalized overlap remains closer to unity over an extended range of separations, indicating a more robust correspondence between the reduced state and the thermal ensemble.

In Figures \ref{overlap8} and \ref{overlap12} (right), we show the ratio of the purities Pur$(\rho)\equiv {\rm Tr} (\rho^2)$ for the subsystem of size $L$ and the thermal system of the same size. One can see that slightly above the string-breaking distance this ratio approaches unity.

\section{Discussion and outlook} \label{sec_discussion}

In this work, we have analyzed the quantum statistical properties of the confining flux tube connecting a static fermion-antifermion pair in the massive Schwinger model. By constructing and diagonalizing the reduced density matrix of the region located between the sources, we were able to probe the microscopic structure of entanglement within the confining string. Our principal finding is that as the separation between the fermion and antifermion approaches the string-breaking distance, the reduced density matrix of the interquark subsystem becomes nearly indistinguishable from a thermal density matrix characterized by an effective temperature determined by the string tension. The overlap between the microscopic and thermal density matrices exhibits a sharp, narrow peak approaching unity at the onset of string breaking, signaling that the confined region undergoes a genuine thermalization transition driven purely by entanglement dynamics.

This emergent thermality arises in the absence of any external reservoir: the subsystem thermalizes because of its entanglement with the remaining degrees of freedom of the field. The process thus provides a concrete realization of how a closed, pure quantum system can exhibit local thermal behavior through entanglement alone. In the present context, the confining flux tube serves as an explicit example of such self-thermalization, with the string-breaking point marking the transition from a coherent, confined state to a mixed ensemble of mesonic fragments that possesses an effective temperature. 

Near the critical separation $d_c$, we observe a level crossing between the two lowest Schmidt eigenstates of the reduced density matrix of half-space. For $d<d_c$, the dominant configuration corresponds to the external charges connected by a confining electric string. As the separation approaches $d_c$, higher Fock components containing additional fermion-antifermion pairs become increasingly important. Beyond the critical distance, the ground state transitions to a configuration in which the external charges are screened by the created fermion and antifermion, forming a two-meson state. The level crossing near $d=d_c$
thus marks the point of maximal entanglement between the string state and the two-meson state, signifying the microscopic origin of the string-breaking transition.

Our results suggest a deep and quantitative connection between confinement, entanglement, and emergent thermality. The flux tube, traditionally viewed as a classical color-electric string, should instead be regarded as a highly entangled quantum state whose reduced subsystems exhibit effective thermodynamic properties. This perspective provides a unifying framework for understanding the apparent “temperature” observed in hadron production and supports the long-standing conjecture that quark confinement and thermalization share a common quantum-informational origin.

Beyond the Schwinger model, the same mechanism is expected to operate in higher-dimensional confining gauge theories, including QCD. In those systems, the formation and subsequent breaking of flux tubes may similarly correspond to entanglement-driven thermalization processes. This correspondence offers a microscopic interpretation of the statistical features of hadronization and opens new avenues for exploring confinement dynamics through measures of quantum entanglement. In the future, it will be interesting to extend the study presented here to QCD; the lattice investigations of entanglement in confining flux tubes of Yang-Mills theory in $(1+1)$ and $(2+1)$ dimensions have already begun \cite{Amorosso:2024glf, Amorosso:2024leg}.
\vskip0.3cm

Our results provide a simple explanation of the ``puzzle of fast thermalization" in high energy collisions -- hadron thermalization may not require any final-state interactions at all, except for the confinement itself. 
\vspace{-0.2cm}
\section*{Acknowledgments}
We are grateful to Adrien Florio, David Frenklakh and Shuzhe Shi for useful discussions and collaboration on related work. We also thank the participants of the QuantHep2025 workshop at Lawrence Berkeley National Lab for insightful comments.  This work was supported in part
by the U.S. Department of Energy, Office of Science, Office of Nuclear Physics, Inqubator for Quantum Simulation (IQuS) under Award Number DOE
(NP) Award DE-SC0020970 (S.G.). This work was also supported by the U.S. Department of Energy, Office of Science, Office of Nuclear Physics, Grants No. DE-FG02-97ER-41014 (UW Nuclear Theory, S.G.), DE-FG88ER41450 (SBU Nuclear Theory, D.K., E.M.) and by the U.S. Department of Energy, Office of Science, National Quantum Information Science Research Centers, Co-design Center for Quantum Advantage (C2QA) under Contract No.DE-SC0012704 (S.G., D.K., E.M.). S.G. was supported in part by a Feodor Lynen Research fellowship of the Alexander von Humboldt foundation. E.M. was supported in part by the Center for Distributed Quantum Processing at Stony Brook University. This work was also supported, in part, by the Department of Physics and the College of Arts and Sciences at the University of Washington.
\bibliography{refs}

@article{Berges:2017zws,
    author = {Berges, J{\"u}rgen and Floerchinger, Stefan and Venugopalan, Raju},
    title = "{Thermal excitation spectrum from entanglement in an expanding quantum string}",
    eprint = "1707.05338",
    archivePrefix = "arXiv",
    primaryClass = "hep-ph",
    doi = "10.1016/j.physletb.2018.01.068",
    journal = "Phys. Lett. B",
    volume = "778",
    pages = "442--446",
    year = "2018"
}

@article{Hentschinski:2024gaa,
    author = "Hentschinski, Martin and Kharzeev, Dmitri E. and Kutak, Krzysztof and Tu, Zhoudunming",
    title = "{QCD evolution of entanglement entropy}",
    eprint = "2408.01259",
    archivePrefix = "arXiv",
    primaryClass = "hep-ph",
    doi = "10.1088/1361-6633/ad910b",
    journal = "Rept. Prog. Phys.",
    volume = "87",
    number = "12",
    pages = "120501",
    year = "2024"
}

@article{Tu:2019ouv,
    author = "Tu, Zhoudunming and Kharzeev, Dmitri E. and Ullrich, Thomas",
    title = "{Einstein-Podolsky-Rosen Paradox and Quantum Entanglement at Subnucleonic Scales}",
    eprint = "1904.11974",
    archivePrefix = "arXiv",
    primaryClass = "hep-ph",
    doi = "10.1103/PhysRevLett.124.062001",
    journal = "Phys. Rev. Lett.",
    volume = "124",
    number = "6",
    pages = "062001",
    year = "2020"
}

@article{Hentschinski:2023izh,
    author = "Hentschinski, Martin and Kharzeev, Dmitri E. and Kutak, Krzysztof and Tu, Zhoudunming",
    title = "{Probing the Onset of Maximal Entanglement inside the Proton in Diffractive Deep Inelastic Scattering}",
    eprint = "2305.03069",
    archivePrefix = "arXiv",
    primaryClass = "hep-ph",
    doi = "10.1103/PhysRevLett.131.241901",
    journal = "Phys. Rev. Lett.",
    volume = "131",
    number = "24",
    pages = "241901",
    year = "2023"
}

@article{Gursoy:2023hge,
    author = {G{\"u}rsoy, Umut and Kharzeev, Dmitri E. and Pedraza, Juan F.},
    title = "{Universal rapidity scaling of entanglement entropy inside hadrons from conformal invariance}",
    eprint = "2306.16145",
    archivePrefix = "arXiv",
    primaryClass = "hep-th",
    reportNumber = "IFT-UAM/CSIC-23-79",
    doi = "10.1103/PhysRevD.110.074008",
    journal = "Phys. Rev. D",
    volume = "110",
    number = "7",
    pages = "074008",
    year = "2024"
}

@article{H1:2020zpd,
    author = "Andreev, V. and others",
    collaboration = "H1",
    title = "{Measurement of charged particle multiplicity distributions in DIS at HERA and its implication to entanglement entropy of partons}",
    eprint = "2011.01812",
    archivePrefix = "arXiv",
    primaryClass = "hep-ex",
    reportNumber = "DESY-20-176",
    doi = "10.1140/epjc/s10052-021-08896-1",
    journal = "Eur. Phys. J. C",
    volume = "81",
    number = "3",
    pages = "212",
    year = "2021"
}

@article{Lipatov:1993yb,
    author = "Lipatov, L. N.",
    title = "{Asymptotic behavior of multicolor QCD at high energies in connection with exactly solvable spin models}",
    eprint = "hep-th/9311037",
    archivePrefix = "arXiv",
    reportNumber = "DFPD-93-TH-70",
    journal = "JETP Lett.",
    volume = "59",
    pages = "596--599",
    year = "1994"
}

@article{Hentschinski:2021aux,
    author = "Hentschinski, Martin and Kutak, Krzysztof",
    title = "{Evidence for the maximally entangled low x proton in Deep Inelastic Scattering from H1 data}",
    eprint = "2110.06156",
    archivePrefix = "arXiv",
    primaryClass = "hep-ph",
    reportNumber = "IFJPAN-IV-2021-16",
    doi = "10.1140/epjc/s10052-022-10056-y",
    journal = "Eur. Phys. J. C",
    volume = "82",
    number = "2",
    pages = "111",
    year = "2022",
    note = "[Erratum: Eur.Phys.J.C 83, 1147 (2023)]"
}

@article{Schlichting:2019abc,
    author = "Schlichting, Soeren and Teaney, Derek",
    title = "{The First fm/c of Heavy-Ion Collisions}",
    eprint = "1908.02113",
    archivePrefix = "arXiv",
    primaryClass = "nucl-th",
    doi = "10.1146/annurev-nucl-101918-023825",
    journal = "Ann. Rev. Nucl. Part. Sci.",
    volume = "69",
    pages = "447--476",
    year = "2019"
}

@article{Kharzeev:2012re,
    author = {Kharzeev, Dmitri E. and Loshaj, Frash{\"e}r},
    title = "{Jet energy loss and fragmentation in heavy ion collisions}",
    eprint = "1212.5857",
    archivePrefix = "arXiv",
    primaryClass = "hep-ph",
    doi = "10.1103/PhysRevD.87.077501",
    journal = "Phys. Rev. D",
    volume = "87",
    number = "7",
    pages = "077501",
    year = "2013"
}

@article{Casher:1974vf,
    author = "Casher, A. and Kogut, John B. and Susskind, Leonard",
    title = "{Vacuum polarization and the absence of free quarks}",
    doi = "10.1103/PhysRevD.10.732",
    journal = "Phys. Rev. D",
    volume = "10",
    pages = "732--745",
    year = "1974"
}

@article{Kharzeev:2006zm,
    author = "Kharzeev, Dmitri and Levin, Eugene and Tuchin, Kirill",
    title = "{Multi-particle production and thermalization in high-energy QCD}",
    eprint = "hep-ph/0602063",
    archivePrefix = "arXiv",
    reportNumber = "BNL-NT-06-8, TAUP-2821-06",
    doi = "10.1103/PhysRevC.75.044903",
    journal = "Phys. Rev. C",
    volume = "75",
    pages = "044903",
    year = "2007"
}

@article{Grieninger:2023ehb,
    author = "Grieninger, Sebastian and Kharzeev, Dmitri E. and Zahed, Ismail",
    title = "{Entanglement in a holographic Schwinger pair with confinement}",
    eprint = "2305.07121",
    archivePrefix = "arXiv",
    primaryClass = "hep-th",
    doi = "10.1103/PhysRevD.108.086030",
    journal = "Phys. Rev. D",
    volume = "108",
    number = "8",
    pages = "086030",
    year = "2023"
}

@article{Florio:2021xvj,
    author = "Florio, Adrien and Kharzeev, Dmitri E.",
    title = "{Gibbs entropy from entanglement in electric quenches}",
    eprint = "2106.00838",
    archivePrefix = "arXiv",
    primaryClass = "hep-th",
    doi = "10.1103/PhysRevD.104.056021",
    journal = "Phys. Rev. D",
    volume = "104",
    number = "5",
    pages = "056021",
    year = "2021"
}

@article{Kharzeev:2005iz,
    author = "Kharzeev, Dmitri and Tuchin, Kirill",
    title = "{From color glass condensate to quark gluon plasma through the event horizon}",
    eprint = "hep-ph/0501234",
    archivePrefix = "arXiv",
    reportNumber = "BNL-NT-05-2",
    doi = "10.1016/j.nuclphysa.2005.03.001",
    journal = "Nucl. Phys. A",
    volume = "753",
    pages = "316--334",
    year = "2005"
}

@article{Grieninger:2023pyb,
    author = "Grieninger, Sebastian and Kharzeev, Dmitri E. and Zahed, Ismail",
    title = "{Entanglement entropy in a time-dependent holographic Schwinger pair creation}",
    eprint = "2310.12042",
    archivePrefix = "arXiv",
    primaryClass = "hep-th",
    doi = "10.1103/PhysRevD.108.126014",
    journal = "Phys. Rev. D",
    volume = "108",
    number = "12",
    pages = "126014",
    year = "2023"
}

@article{Castorina:2007eb,
    author = "Castorina, P. and Kharzeev, D. and Satz, H.",
    title = "{Thermal Hadronization and Hawking-Unruh Radiation in QCD}",
    eprint = "0704.1426",
    archivePrefix = "arXiv",
    primaryClass = "hep-ph",
    reportNumber = "BNL-NT-07-18, BI-TP-2007-06",
    doi = "10.1140/epjc/s10052-007-0368-6",
    journal = "Eur. Phys. J. C",
    volume = "52",
    pages = "187--201",
    year = "2007"
}

@article{Andronic_2009,
   title={Thermal description of hadron production in $e^+e^-$ collisions revisited},
   volume={675},
   ISSN={0370-2693},
   url={http://dx.doi.org/10.1016/j.physletb.2009.04.024},
   DOI={10.1016/j.physletb.2009.04.024},
   number={3–4},
   journal={Physics Letters B},
   publisher={Elsevier BV},
   author={Andronic, A. and Beutler, F. and Braun-Munzinger, P. and Redlich, K. and Stachel, J.},
   year={2009},
   month=may, pages={312–318} }

@unpublished{Grieninger:2025wxg,
    author = "Grieninger, Sebastian and Hao, Kun and Kharzeev, Dmitri E. and Korepin, Vladimir",
    title = "{Small $x$ behavior in QCD from maximal entanglement and conformal invariance}",
    eprint = "2508.21643",
    archivePrefix = "arXiv",
    primaryClass = "hep-ph",
    year = "2025"
}

@unpublished{Xu:2025abo,
    author = "Xu, Kaidi and Borla, Umberto and Moroz, Sergej and Halimeh, Jad C.",
    title = "{String Breaking Dynamics and Glueball Formation in a $2+1$D Lattice Gauge Theory}",
    eprint = "2507.01950",
    archivePrefix = "arXiv",
    primaryClass = "hep-lat",
    month = "7",
    year = "2025"
}

@unpublished{Crippa:2024hso,
    author = "Crippa, Arianna and Jansen, Karl and Rinaldi, Enrico",
    title = "{Analysis of the confinement string in (2 + 1)-dimensional Quantum Electrodynamics with a trapped-ion quantum computer}",
    eprint = "2411.05628",
    archivePrefix = "arXiv",
    primaryClass = "hep-lat",
    month = "11",
    year = "2024"
}

@unpublished{De:2024smi,
    author = "De, Arinjoy and others",
    title = "{Observation of string-breaking dynamics in a quantum simulator}",
    eprint = "2410.13815",
    archivePrefix = "arXiv",
    primaryClass = "quant-ph",
    month = "10",
    year = "2024"
}

@unpublished{Artiaco:2025qqq,
    author = "Artiaco, Claudia and Barata, Jo{\~a}o and Rico, Enrique",
    title = "{Out-of-Equilibrium Dynamics in a U(1) Lattice Gauge Theory via Local Information Flows: Scattering and String Breaking}",
    eprint = "2510.16101",
    archivePrefix = "arXiv",
    primaryClass = "quant-ph",
    reportNumber = "CERN-TH-2025-191",
    month = "10",
    year = "2025"
}

@article{Buyens:2015tea,
    author = "Buyens, Boye and Haegeman, Jutho and Verschelde, Henri and Verstraete, Frank and Van Acoleyen, Karel",
    title = "{Confinement and string breaking for QED$_2$ in the Hamiltonian picture}",
    eprint = "1509.00246",
    archivePrefix = "arXiv",
    primaryClass = "hep-lat",
    doi = "10.1103/PhysRevX.6.041040",
    journal = "Phys. Rev. X",
    volume = "6",
    number = "4",
    pages = "041040",
    year = "2016"
}

@unpublished{Surace:2024bht,
    author = "Surace, Federica Maria and others",
    title = "{String-Breaking Dynamics in Quantum Adiabatic and Diabatic Processes}",
    eprint = "2411.10652",
    archivePrefix = "arXiv",
    primaryClass = "quant-ph",
    month = "11",
    year = "2024"
}

@article{Verdel:2023mmp,
    author = "Verdel, Roberto and Zhu, Guo-Yi and Heyl, Markus",
    title = "{Dynamical Localization Transition of String Breaking in Quantum Spin Chains}",
    eprint = "2304.12957",
    archivePrefix = "arXiv",
    primaryClass = "cond-mat.str-el",
    doi = "10.1103/PhysRevLett.131.230402",
    journal = "Phys. Rev. Lett.",
    volume = "131",
    number = "23",
    pages = "230402",
    year = "2023"
}

@article{Mallick:2024slg,
    author = "Mallick, Arindam and Lewenstein, Maciej and Zakrzewski, Jakub and P{\l}odzie{\'n}, Marcin",
    title = "{String-breaking dynamics in an Ising chain with local vibrations}",
    eprint = "2501.00604",
    archivePrefix = "arXiv",
    primaryClass = "quant-ph",
    doi = "10.1103/mdcm-5w9k",
    journal = "Phys. Rev. B",
    volume = "112",
    number = "2",
    pages = "024311",
    year = "2025"
}

@article{Verdel:2019chj,
    author = "Verdel, Roberto and Liu, Fangli and Whitsitt, Seth and Gorshkov, Alexey V. and Heyl, Markus",
    title = "{Real-time dynamics of string breaking in quantum spin chains}",
    eprint = "1911.11382",
    archivePrefix = "arXiv",
    primaryClass = "cond-mat.stat-mech",
    doi = "10.1103/PhysRevB.102.014308",
    journal = "Phys. Rev. B",
    volume = "102",
    number = "1",
    pages = "014308",
    year = "2020"
}

@unpublished{Borla:2025gfs,
    author = "Borla, Umberto and Osborne, Jesse J. and Moroz, Sergej and Halimeh, Jad C.",
    title = "{String Breaking in a $2+1$D $\mathbb{Z}_2$ Lattice Gauge Theory}",
    eprint = "2501.17929",
    archivePrefix = "arXiv",
    primaryClass = "quant-ph",
    month = "1",
    year = "2025"
}

@article{Liu:2024lut,
    author = "Liu, Ying and Zhang, Wei-Yong and Zhu, Zi-Hang and He, Ming-Gen and Yuan, Zhen-Sheng and Pan, Jian-Wei",
    title = "{String-Breaking Mechanism in a Lattice Schwinger Model Simulator}",
    eprint = "2411.15443",
    archivePrefix = "arXiv",
    primaryClass = "cond-mat.quant-gas",
    doi = "10.1103/mwy1-v9hk",
    journal = "Phys. Rev. Lett.",
    volume = "135",
    number = "10",
    pages = "101902",
    year = "2025"
}

@unpublished{DiMarcantonio:2025cmf,
    author = "Di Marcantonio, Francesco and Pradhan, Sunny and Vallecorsa, Sofia and Ba{\~n}uls, Mari Carmen and Ortega, Enrique Rico",
    title = "{Roughening and dynamics of an electric flux string in a (2+1)D lattice gauge theory}",
    eprint = "2505.23853",
    archivePrefix = "arXiv",
    primaryClass = "hep-lat",
    reportNumber = "CERN-TH-2025-105",
    month = "5",
    year = "2025"
}

@unpublished{Cataldi:2025cyo,
    author = "Cataldi, Giovanni and Orlando, Simone and Halimeh, Jad C.",
    title = "{Real-Time String Dynamics in a $2+1$D Non-Abelian Lattice Gauge Theory: String Breaking, Glueball Formation, Baryon Blockade, and Tension Reduction}",
    eprint = "2509.08868",
    archivePrefix = "arXiv",
    primaryClass = "hep-lat",
    month = "9",
    year = "2025"
}

@unpublished{Alexandrou:2025vaj,
    author = {Alexandrou, Constantia and Athenodorou, Andreas and Blekos, Kostas and Polykratis, Georgios and K{\"u}hn, Stefan},
    title = "{Realizing string breaking dynamics in a $Z_2$ lattice gauge theory on quantum hardware}",
    eprint = "2504.13760",
    archivePrefix = "arXiv",
    primaryClass = "hep-lat",
    month = "4",
    year = "2025"
}

@article{Kharzeev_2017,
   title={Deep inelastic scattering as a probe of entanglement},
   volume={95},
   ISSN={2470-0029},
   url={http://dx.doi.org/10.1103/PhysRevD.95.114008},
   DOI={10.1103/physrevd.95.114008},
   number={11},
   journal={Physical Review D},
   publisher={American Physical Society (APS)},
   author={Kharzeev, Dmitri E. and Levin, Eugene M.},
   year={2017},
   month=jun }

@article{Datta:2024hpn,
    author = {Datta, Jaydeep and Deshpande, Abhay and Kharzeev, Dmitri E. and Na{\"\i}m, Charles Joseph and Tu, Zhoudunming},
    title = "{Entanglement as a Probe of Hadronization}",
    eprint = "2410.22331",
    archivePrefix = "arXiv",
    primaryClass = "hep-ph",
    doi = "10.1103/PhysRevLett.134.111902",
    journal = "Phys. Rev. Lett.",
    volume = "134",
    number = "11",
    pages = "111902",
    year = "2025"
}

@article{Gonzalez-Cuadra:2024xul,
    author = "Gonzalez-Cuadra, Daniel and others",
    title = "{Observation of string breaking on a (2 + 1)D Rydberg quantum simulator}",
    eprint = "2410.16558",
    archivePrefix = "arXiv",
    primaryClass = "quant-ph",
    doi = "10.1038/s41586-025-09051-6",
    journal = "Nature",
    volume = "642",
    number = "8067",
    pages = "321--326",
    year = "2025"
}

@article{Cochran:2024rwe,
    author = "Cochran, Tyler A. and others",
    title = "{Visualizing dynamics of charges and strings in (2 + 1)D lattice gauge theories}",
    eprint = "2409.17142",
    archivePrefix = "arXiv",
    primaryClass = "quant-ph",
    doi = "10.1038/s41586-025-08999-9",
    journal = "Nature",
    volume = "642",
    number = "8067",
    pages = "315--320",
    year = "2025"
}

@article{Sonner:2013mba,
    author = "Sonner, Julian",
    title = "{Holographic Schwinger Effect and the Geometry of Entanglement}",
    eprint = "1307.6850",
    archivePrefix = "arXiv",
    primaryClass = "hep-th",
    reportNumber = "MIT-CTP-4483",
    doi = "10.1103/PhysRevLett.111.211603",
    journal = "Phys. Rev. Lett.",
    volume = "111",
    number = "21",
    pages = "211603",
    year = "2013"
}

@article{Jensen:2013ora,
    author = "Jensen, Kristan and Karch, Andreas",
    title = "{Holographic Dual of an Einstein-Podolsky-Rosen Pair has a Wormhole}",
    eprint = "1307.1132",
    archivePrefix = "arXiv",
    primaryClass = "hep-th",
    doi = "10.1103/PhysRevLett.111.211602",
    journal = "Phys. Rev. Lett.",
    volume = "111",
    number = "21",
    pages = "211602",
    year = "2013"
}

@article{Florio:2025hoc,
title = {Thermalization from quantum entanglement: Jet simulations in the massive Schwinger model},
  author = {Florio, Adrien and Frenklakh, David and Grieninger, Sebastian and Kharzeev, Dmitri E. and Palermo, Andrea and Shi, Shuzhe},
  journal = {Phys. Rev. D},
  volume = {112},
  issue = {9},
  pages = {094502},
  numpages = {20},
  year = {2025},
  month = {Nov},
  publisher = {American Physical Society},
  doi = {10.1103/sgrx-jpp9},
  url = {https://link.aps.org/doi/10.1103/sgrx-jpp9}
}

@article{Kharzeev:2021nzh,
    author = "Kharzeev, Dmitri E.",
    title = "{Quantum information approach to high energy interactions}",
    eprint = "2108.08792",
    archivePrefix = "arXiv",
    primaryClass = "hep-ph",
    doi = "10.1098/rsta.2021.0063",
    journal = "Phil. Trans. A. Math. Phys. Eng. Sci.",
    volume = "380",
    number = "2216",
    pages = "20210063",
    year = "2021"
}

@article{Kharzeev:2021yyf,
    author = "Kharzeev, Dmitri E. and Levin, Eugene",
    title = "{Deep inelastic scattering as a probe of entanglement: Confronting experimental data}",
    eprint = "2102.09773",
    archivePrefix = "arXiv",
    primaryClass = "hep-ph",
    doi = "10.1103/PhysRevD.104.L031503",
    journal = "Phys. Rev. D",
    volume = "104",
    number = "3",
    pages = "L031503",
    year = "2021"
}

@article{Becattini:1997uf,
    author = "Becattini, F.",
    editor = "Panagiotou, A. D.",
    title = "{Thermal hadron production in high-energy collisions}",
    eprint = "hep-ph/9708248",
    archivePrefix = "arXiv",
    reportNumber = "DFF-284-07-1997",
    doi = "10.1088/0954-3899/23/12/017",
    journal = "J. Phys. G",
    volume = "23",
    pages = "1933--1940",
    year = "1997"
}

@article{Schwinger2,
  title = {Gauge Invariance and Mass. II},
  author = {Schwinger, Julian},
  journal = {Phys. Rev.},
  volume = {128},
  issue = {5},
  pages = {2425--2429},
  numpages = {0},
  year = {1962},
  month = {Dec},
  publisher = {American Physical Society},
  doi = {10.1103/PhysRev.128.2425},
  url = {https://link.aps.org/doi/10.1103/PhysRev.128.2425}
}

@article{kogutsuss,
  title = {Hamiltonian formulation of Wilson's lattice gauge theories},
  author = {Kogut, John and Susskind, Leonard},
  journal = {Phys. Rev. D},
  volume = {11},
  issue = {2},
  pages = {395--408},
  numpages = {0},
  year = {1975},
  month = {Jan},
  publisher = {American Physical Society},
  doi = {10.1103/PhysRevD.11.395},
  url = {https://link.aps.org/doi/10.1103/PhysRevD.11.395}
}

@article{PhysRevD.16.3031,
  title = {Lattice fermions},
  author = {Susskind, Leonard},
  journal = {Phys. Rev. D},
  volume = {16},
  issue = {10},
  pages = {3031--3039},
  numpages = {0},
  year = {1977},
  month = {Nov},
  publisher = {American Physical Society},
  doi = {10.1103/PhysRevD.16.3031},
  url = {https://link.aps.org/doi/10.1103/PhysRevD.16.3031}
}

@article{Amorosso:2024glf,
    author = "Amorosso, Rocco and Syritsyn, Sergey and Venugopalan, Raju",
    title = "{Entanglement entropy of a color flux tube in (1+1)D Yang{\textendash}Mills theory}",
    eprint = "2411.12818",
    archivePrefix = "arXiv",
    primaryClass = "hep-lat",
    doi = "10.1016/j.physletb.2025.139806",
    journal = "Phys. Lett. B",
    volume = "868",
    pages = "139806",
    year = "2025"
}

@article{Amorosso:2024leg,
    author = "Amorosso, Rocco and Syritsyn, Sergey and Venugopalan, Raju",
    title = "{Entanglement entropy of a color flux tube in (2+1)D Yang-Mills theory}",
    eprint = "2410.00112",
    archivePrefix = "arXiv",
    primaryClass = "hep-lat",
    doi = "10.1007/JHEP12(2024)177",
    journal = "JHEP",
    volume = "12",
    pages = "177",
    year = "2024"
}

@article{Florio:2023dke,
    author = "Florio, Adrien and Frenklakh, David and Ikeda, Kazuki and Kharzeev, Dmitri and Korepin, Vladimir and Shi, Shuzhe and Yu, Kwangmin",
    title = "{Real-Time Nonperturbative Dynamics of Jet Production in Schwinger Model: Quantum Entanglement and Vacuum Modification}",
    eprint = "2301.11991",
    archivePrefix = "arXiv",
    primaryClass = "hep-ph",
    doi = "10.1103/PhysRevLett.131.021902",
    journal = "Phys. Rev. Lett.",
    volume = "131",
    number = "2",
    pages = "021902",
    year = "2023"
}

@article{Florio:2024aix,
    author = "Florio, Adrien and Frenklakh, David and Ikeda, Kazuki and Kharzeev, Dmitri E. and Korepin, Vladimir and Shi, Shuzhe and Yu, Kwangmin",
    title = "{Quantum real-time evolution of entanglement and hadronization in jet production: Lessons from the massive Schwinger model}",
    eprint = "2404.00087",
    archivePrefix = "arXiv",
    primaryClass = "hep-ph",
    doi = "10.1103/PhysRevD.110.094029",
    journal = "Phys. Rev. D",
    volume = "110",
    number = "9",
    pages = "094029",
    year = "2024"
}

@misc{bezanson2015juliafreshapproachnumerical,
      title={Julia: A Fresh Approach to Numerical Computing}, 
      author={Jeff Bezanson and Alan Edelman and Stefan Karpinski and Viral B. Shah},
      year={2015},
      eprint={1411.1607},
      archivePrefix={arXiv},
      primaryClass={cs.MS},
      url={https://arxiv.org/abs/1411.1607}, 
}

@article{Fishman_2022,
   title={The ITensor Software Library for Tensor Network Calculations},
   url={http://dx.doi.org/10.21468/SciPostPhysCodeb.4},
   DOI={10.21468/scipostphyscodeb.4},
   journal={SciPost Physics Codebases},
   publisher={Stichting SciPost},
   author={Fishman, Matthew and White, Steven and Stoudenmire, Edwin},
   year={2022},
   month=aug }

@unpublished{Barata:2025hgx,
    author = "Barata, Jo{\~a}o and Rico, Enrique",
    title = "{Real-time simulation of jet energy loss and entropy production in high-energy scattering with matter}",
    eprint = "2502.17558",
    archivePrefix = "arXiv",
    primaryClass = "hep-ph",
    reportNumber = "CERN-TH-2025-019",
    month = "2",
    year = "2025"
}

@article{Janik:2025bbz,
    author = "Janik, Romuald A. and Nowak, Maciej A. and Rams, Marek M. and Zahed, Ismail",
    title = "{Emergent Nonthermal Fluid from Jets in the Massive Schwinger Model Using Tensor Networks}",
    eprint = "2502.12901",
    archivePrefix = "arXiv",
    primaryClass = "hep-ph",
    doi = "10.1103/gvr2-gqys",
    journal = "Phys. Rev. Lett.",
    volume = "135",
    number = "21",
    pages = "211903",
    year = "2025"
}

@article{Turkeshi_2020,
   title={Entanglement equipartition in critical random spin chains},
   volume={102},
   ISSN={2469-9969},
   url={http://dx.doi.org/10.1103/PhysRevB.102.014455},
   DOI={10.1103/physrevb.102.014455},
   number={1},
   journal={Physical Review B},
   publisher={American Physical Society (APS)},
   author={Turkeshi, Xhek and Ruggiero, Paola and Alba, Vincenzo and Calabrese, Pasquale},
   year={2020},
   month=jul }

@article{Asadi:2022vbl,
    author = "Asadi, Pouya and Vaidya, Varun",
    title = "{Quantum entanglement and the thermal hadron}",
    eprint = "2211.14333",
    archivePrefix = "arXiv",
    primaryClass = "nucl-th",
    doi = "10.1103/PhysRevD.107.054028",
    journal = "Phys. Rev. D",
    volume = "107",
    number = "5",
    pages = "054028",
    year = "2023"
}

@article{Asadi:2023bat,
    author = "Asadi, Pouya and Vaidya, Varun",
    title = "{1+1D hadrons minimize their biparton Renyi free energy}",
    eprint = "2301.03611",
    archivePrefix = "arXiv",
    primaryClass = "hep-th",
    doi = "10.1103/PhysRevD.108.014036",
    journal = "Phys. Rev. D",
    volume = "108",
    number = "1",
    pages = "014036",
    year = "2023"
}

\appendix
\bigskip

\section{Symmetry resolved entanglement entropy}\label{app:sre}
To quantify the entanglement in the ground state, we consider partitions of the full Hilbert space $\mathcal{H}= \mathcal{H}_A \otimes \mathcal{H}_B$ into two complementary subsystems. Depending on the observable of interest, $\mathcal{H}_A$ may represent either one half of the lattice or a subsystem of length $L$ centered in the middle of the lattice (see Fig. \ref{setup_fig}), while $\mathcal{H}_B$ corresponds to the remaining sites of the system.

Given the ground state $\ket{\Psi_0}$ of the full system, the reduced density matrix of subsystem $A$ is obtained by tracing out its complement:
\begin{equation}
    \rho_A = \textrm{Tr}_B(\ket{\Psi_0}\bra{\Psi_0}).
\end{equation}
Diagonalizing $\rho_A$ gives the \emph{entanglement spectrum}, defined as the set of eigenvalues $\{\lambda_i\}$ of $\rho_A$. Equivalently, the ground state admit a Schmidt decomposition
\begin{equation}
    \ket{\Psi_0} = \sum_{i=1}^{\chi} \sqrt{\lambda_i} \ket{\psi_i^ A} \otimes \ket{\psi_i^ B}
\end{equation}
where $\lambda_i\geq0$, $\sum_i\lambda_i=1$, and $\chi$ is the Schmidt rank. These coefficients are the eigenvalues of both $\rho_A$ and $\rho_B$. From them, the EE follows as
\begin{equation}
    S_{EE}=-\sum_i \lambda_i \ln{\lambda_i}.
\end{equation}
While $S_{EE}$ provides a single quantitative measure of entanglement, the entanglement spectrum carries the full structure information of quantum correlations within the subsystem, allowing us to identify near-degeneracies, level of rearrangements, and mixing of the density matrix.
\bigskip

In the presence of a global $U(1)$ symmetry associated with fermion number conservation, the reduced density matrix of a subsystem commutes with the local charge operator $Q_A$, $[\rho_A, Q_A]=0$. As a result, $\rho_A$ is block-diagonal in the eigenvalues $Q_A$ of the conserved charge and can be decomposed as \cite{Turkeshi_2020}
\begin{align}
&\rho_A = \bigoplus_Q p_Q\,\rho_Q, \quad  \textrm{with}
\\ 
&p_Q = \mathrm{Tr}\!\left(\Pi_Q \rho_A\right),
\quad \text{and} \quad
\rho_Q = \frac{\Pi_Q \rho_A \Pi_Q}{p_Q} ,
\end{align}
where $\Pi_Q$ is the projector onto the subspace with total charge $Q$. The probability $p(Q)$ quantifies the fluctuations of the conserved charge within the subsystem, while $\rho_A$ is the normalized reduced density restricted to that charge sector.

The von Neumann entanglement entropy naturally separates into a 
\emph{number (fluctuation) entropy} $S_Q$ and a \emph{configurational entropy} $S_{\mathrm{conf}}$
\begin{align}
S_{EE}(\rho_A) &= S_{\textrm{fluct}} + S_{\mathrm{conf}},
\qquad \\
S_{\textrm{fluct}} &= -\sum_Q p_Q\,\ln p_Q,
\qquad \\
S_{\mathrm{conf}} &= \sum_Q p_Q\,S^Q_{EE} ,\label{eq:conf}
\end{align}
with
\begin{equation}
S^Q_{EE} = -\,\mathrm{Tr}\!\left[\rho_Q\ln\rho_Q\right].
\end{equation}
\\
The quantity $S^Q_{EE}$ defines the \emph{symmetry-resolved entanglement entropy}, which measures the entanglement within each charge sector $Q$.

\begin{figure}[ht!]
     \centering
     \includegraphics[width=0.8\linewidth]{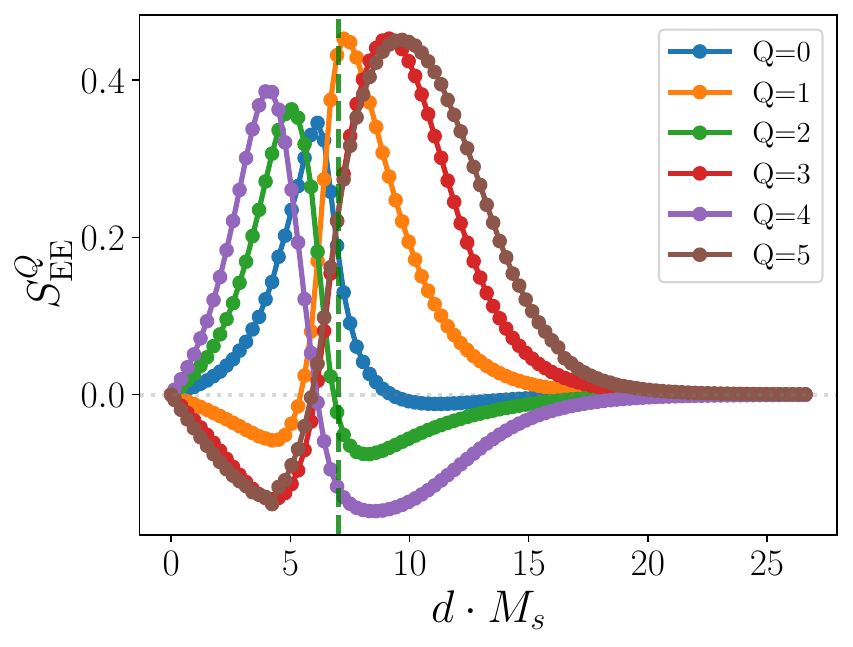}       \includegraphics[width=0.8\linewidth]{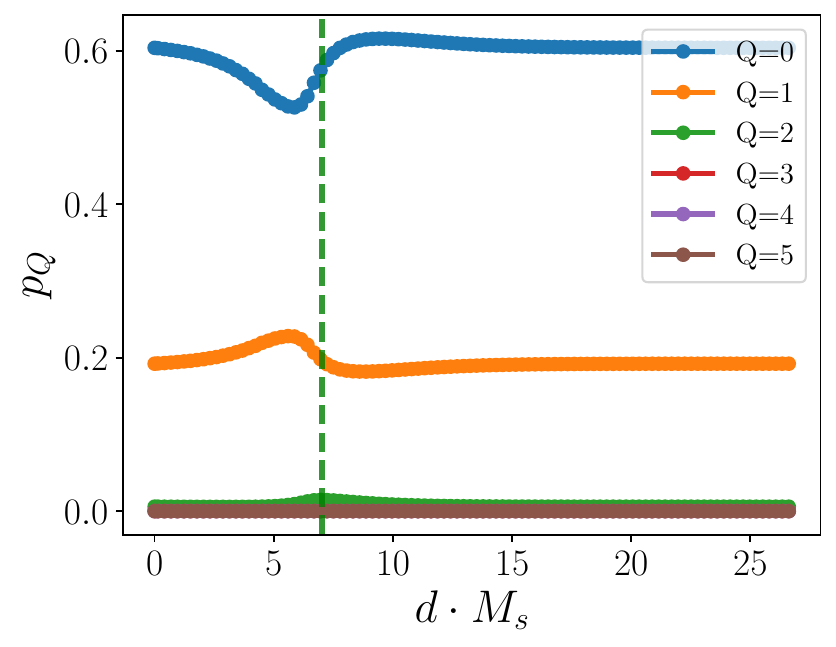}
     \caption{Symmetry-resolved entanglement entropy $S^Q_{EE}$ (\textbf{top}) and the corresponding sector weights (\textbf{bottom}) for a centered subsystem of size $L=32$.}
     \label{symre}
\end{figure}

Figure \ref{symre} (top) shows the behavior of the symmetry-resolved entanglement entropy in different charge sectors as the separation between the external charges increases. This analysis is performed for a centered subsystem of size $L=32$ in the middle of the lattice, where $Q$ is the net charge enclosed within the subsystem. Fig. \ref{symre} (bottom) shows the weights of different charge sectors. Together with the symmetry resolved EE, the weights  define the configuration entropy~\eqref{eq:conf}.

\end{document}